\documentclass{IEEEtran}
\usepackage{cite}
\usepackage{amsmath,amssymb,amsfonts,nccmath}
\usepackage{algorithmic}
\usepackage{graphicx}
\usepackage{textcomp}
\usepackage{xcolor}
\usepackage{upgreek}
\usepackage{cases}
\usepackage{scalerel}
\usepackage[normalem]{ulem}
\def\BibTeX{{\rm B\kern-.05em{\sc i\kern-.025em b}\kern-.08em
    T\kern-.1667em\lower.7ex\hbox{E}\kern-.125emX}}
\begin{document}
\title{The Evaluation of a Novel Asymptotic Solution to the Sommerfeld Radiation Problem using an Efficient Method for the Calculation of Sommerfeld Integrals in the Spectral Domain}
\author{Sotiris Bourgiotis, Panayiotis Frangos, Seil Sautbekov and Mustakhim Pshikov
\thanks{S. Bourgiotis and P. Frangos are with the School of Electrical and Computer Engineering, National Technical University of Athens, Athens 15773, Greece (e-mail: sbourgiotis@mail.ntua.gr ; \mbox{pfrangos}@central.ntua.gr)}
\thanks{S. Sautbekov and M. Pshikov are with the department of Physics and Technology, Al-Farabi Kazakh National University, 71 Al-Farabi Ave., Almaty, Kazakhstan (e-mail: sautbek@mail.ru ; Mustahim64@mail.ru).}}

\maketitle

\begin{abstract}
In this work, a recently developed novel solution of the famous "Sommerfeld Radiation Problem" is revisited. The solution is based on an analysis performed entirely in the spectral domain, through which a compact asymptotic formula describes the behavior of the EM field, which emanates from a vertical Hertzian radiating dipole, located above flat, lossy ground. The paper is divided into two parts. First, we demonstrate an efficient technique for the accurate numeric calculation of the well--known Sommerfeld integrals, required for the evaluation of the field. The results are compared against alternative calculation approaches and validated with the corresponding Norton figures for the Surface Wave. Then, in the second part, we briefly introduce the asymptotic solution of interest and investigate its performance; we contrast the solution versus the accurate numerical evaluation for the total received EM field and also with a more basic asymptotic solution to the given problem, obtained via the application of the Stationary Phase Method (SPM). Simulations for various frequencies, distances, altitudes and ground characteristics are illustrated and inferences for the applicability of the solution are made. Finally, special cases, leading to analytic field expressions, close as well as far from the interface, are examined.
\end{abstract}

\begin{IEEEkeywords}
Asymptotic Solution, Hertzian Dipole, Numerical Integration, Sommerfeld Radiation Problem, Surface Wave.
\end{IEEEkeywords}
\vspace{-3mm}
\section{Introduction}
\label{sec:introduction}
\IEEEPARstart{T}{he} Sommerfeld radiation problem is a classical problem in the area of Electromagnetic (EM) waves propagation above the terrain. The original Sommerfeld solution to this problem is provided in the spatial domain as an integral expression, utilizing the so-called ''Hertz potentials'', but it does not end up into closed-form analytic formulas\cite{Sommerfeld1909, Sommerfeld1926, Wait1998, King1969, Zenneck1907, Sarkar2012, Bladel2007, Banos1966, Tyras1969, Rahmat1981, Collin2004, Michalski1985, Pelosi2010}. Moreover, for the calculation of the EM field, it is necessary to evaluate the derivatives of those potentials, which poses an extra accuracy issue. Subsequently, K. A. Norton focused on the engineering application of the problem and provided approximate solutions, represented by long algebraic expressions, describing concepts, such as the propagating \emph{Surface Wave} and its associated ''attenuation coefficient''\cite{Norton1936, Norton1937}.


\subsection{Previous Contribution of our Research Group}
\label{subsec:history}
The problem may also be tackled in the spectral domain. Particularly, in\cite{CEMA2010, Sautbekov2011}, we derived the fundamental integral representations for the received EM field, by means of a generalized solution to the respective Maxwell equations boundary value problem. Working in this domain has the advantage that no Hertz potentials and their subsequent differentiation, are required for the evaluation of the fields.

In\cite{Ioannidi2014}, the Stationary Phase Method (SPM)\cite{Balanis1997, Moschovitis2010, Moschovitis2010PIER} was applied, to the general integral expressions of the EM field and the well-known analytic formulas for the \emph{Space Wave}, defined as the complex summation of the \emph{Line of Sight} (LOS) field and a portion of the field emanating from the dipole's image point (also called the \emph{Reflected Field}), were obtained as the high frequency asymptotic solution to the complete problem. In\cite{Bourgiotis2014, Bourgiotis2015, Chrysostomou2016}, we focused on the numerical evaluation of the field's integral expressions and how they compare with their respective high frequency approximations. It was revealed that accurately evaluating the Sommerfeld Integrals in the spectral domain is also not a trivial task. The result is sensitive on the position of the singular points in relation to the integration path, an issue that has been also a major problem and a matter of debate in various related research works\cite{Sarkar2012}.

Then in\cite{Sautbekov2018}, the mathematical formulation of the problem was redefined, for the case where $\upsigma \gg \upomega\upvarepsilon_{0}$, which is valid for many practical frequencies of interest in terrestrial communications. As shown there, a special contour integral, called ''Etalon Integral'',  was used to deform the original contour of integration, through the application of the Saddle Point Method (SDP). This ''Etalon Integral'' can be expressed in terms of Fresnel Integrals and has interesting properties, which can reduce the problem related to the vicinity of the saddle point to the pole point\cite{Weinstein1969, Fock1945, Fock1946, Leontovic1946, Pedlosky2003, Sautbekov2010}. The result was a compact asymptotic formula that better expresses the variation of the field in the high frequency regime. Moreover, using the small and large argument approximations, associated with the Fresnel Integrals\cite{DLMF}, pure analytic expressions were extracted, describing the behavior of the EM field close, as well as far away from the ground interface.

\subsection{Scope of this Research}
\label{subsec:scope}
\begin{figure}[!t]
\centerline{\includegraphics[width=\columnwidth, height=0.70\columnwidth]{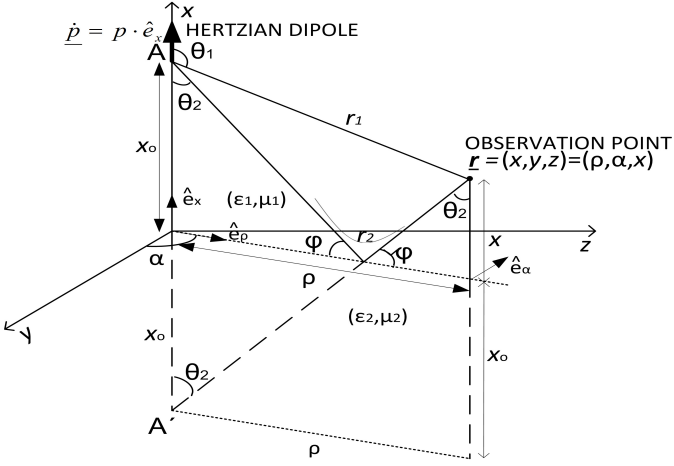}}
\caption{Hertzian Dipole above infinite, planar interface. Point A$^{^\prime}$ is the image of the source A with respect to the ground (\textit{yz}-plane), r\textsubscript{1} is the distance between the source and the observation point, r\textsubscript{2} is the distance between the image of the source and the observation point, $\uptheta_{2}$ is the ``angle of incidence'' at the so-called ``specular point'', which is the point of intersection of the ground (\textit{yz}-plane) with the line connecting the image point and the observation point, and finally, $\upvarphi = \pi/2 - \uptheta_{2}$ is the so-called ``grazing angle''.}
\label{fig:1}
\end{figure}

As mentioned in\cite{Bourgiotis2015, Chrysostomou2016}, the accurate evaluation of the Sommerfeld integral expressions is not a straightforward task and this is due to both the presence of singularities along the integration path as well as the particular complex and rapidly oscillating nature of the integrands. For that reason various specialized commercial software have been used for obtaining adequate results, for example the AWAS tool used by Sarkar \emph{et al.} in\cite{Sarkar2012}. In this paper we show that using an appropriate variable transformation, it is possible to convert the generalized integrals of\cite{Ioannidi2014} into fast converging formulas, which are rather suitable for numerical calculation, using standard numerical integration techniques. Particularly, the integral expression describing the received EM field, is broken down into two terms; one relatively easily computed definite integral, of finite integation range and another integral of semi-infinite range. However, the latter integrand proves to be a rapidly decaying exponential function, resulting in very fast convergence times. Comparisons against the numerical results published in\cite{Bourgiotis2015, Chrysostomou2016} demonstrate the advantage of the method. Also a validation against Norton's figures, associated with the well-known surface wave\cite{Norton1936, Norton1937}, is exhibited. Preliminary results have already been demonstrated in\cite{Bourgiotis2018}.

The analysis in\cite{Sautbekov2018} was constrained to the pure mathematical formulation of the problem. Simulations and related figures are still pending for validating the method. Hence, in the second part of the paper, extensive simulations are demonstrated for this purpose. We compare the method's performance against the more basic asymptotic solution of\cite{Ioannidi2014}. The latter is based on the application of the SPM method and essentially yields the geometric optics field. Comparissons are made against the accurate numerical integrations results, which are used as the baseline. Moreover, an investigation of the above mentioned analytic formulas, which according to the analysis in\cite{Sautbekov2018}, should reflect the field behavior close to as well as far away from the ground level, is made. Provided the required conditions are satisfied, our simulations validate the analytic expressions.

\subsection{Problem Geometry}
\label{subsec:geometry}
The geometry of the problem, as given in\cite{Sautbekov2018} and also briefly described here for ease of reference, is shown in Fig.~\ref{fig:1}. A vertical small (Hertzian) dipole, characterized by dipole moment $\underline{\dot{p}}=p\cdot\hat{e}_{x}, ~p\text{=const}$, is directed to the positive x axis, at altitude {\Large x\textsubscript{o}} above infinite, flat and lossy ground. The dipole radiates time-harmonic electromagnetic (EM) waves at angular frequency $\upomega = 2\pi f$ ($e^{-i\upomega t}$ time dependence is assumed). The relative complex permittivity of the ground is:~ $\upvarepsilon_{r}^{'} = \upvarepsilon^{'} / \upvarepsilon_{0} = \upvarepsilon_{r} + i \upsigma / \upomega \upvarepsilon_{0}$, where $\upsigma$ is the ground conductivity, $f$ the carrier frequency and $\upvarepsilon_{0} = 8.854\times10^{-12}$F/m is the permitivity in vacuum or air. The goal is to evaluate the received EM field at an arbitrary observation point above the ground level, namely at  point {\large(x,y,z)}, shown in Fig.~\ref{fig:1}.

\subsection{Structure of the Article}
\label{subsec:structure}
In what follows, Section \ref{sec:formulation} recaps the fundamentals expressions for the EM field in the spectral domain, the issues associated with their numerical calculation and demonstrates how a simple variable transformation may lead to fast converging integral formulas, suitable for evaluation in the computer. Through various simulation results, we illustrate the advantages and validate the accuracy of the redefined expressions. Then in Section \ref{sec:asymptotic_eval}, we give a brief overview of the asymptotic solution of\cite{Sautbekov2018} and through an extended set of simulations -- comparisons we demonstrate its efficiency. Moreover, a discussion regarding the applicability of the closed-form formulas, also predicted by\cite{Sautbekov2018}, is given. Finally, in Section \ref{sec:conclusion}  we summarize on the major findings and we make a brief discussion on potential extensions. The whole analysis is given for the electric field. Expressions for the magnetic field are derived similarly or by suitable use of the duality principle.

\section{Efficient Formulation for the EM Field Integral Expressions in the Spectral Domain}
\label{sec:formulation}

\subsection{Original Integral Expressions}
\label{subsec:orig_formulas}
According to the analysis of\cite{Ioannidi2014}, performed in the spectral domain, the electric field at the observation point of Fig. \ref{fig:1} is given by the following integral expression,
\begin{IEEEeqnarray}{rCl}
\IEEEeqnarraymulticol{3}{c}{
\underline{E} = \underline{E}^{LOS} + \underline{E}^{R}}\nonumber \vspace{1mm}\\
& = & -\dfrac{ip}{8\uppi\upvarepsilon_{0}\upvarepsilon_{1}}\left[\int_{-\infty}^{\infty}\underline{\mathrm{f_1}}(k_{\uprho})\mathrm{d}k_{\uprho} +\int_{-\infty}^{\infty}\underline{\mathrm{f_2}}(k_{\uprho})\mathrm{d}k_{\uprho}\right],\label{eq:1}
\end{IEEEeqnarray}
where $\underline{E}^{LOS}$ denotes the direct or LOS field, $\underline{E}^{R}$ is for the field scattered by the flat and lossy ground and the vector functions $\underline{\mathrm{f_1}}(k_{\uprho})$ and $\underline{\mathrm{f_2}}(k_{\uprho})$ are given by 
\begin{IEEEeqnarray}{rCl}
\underline{\mathrm{f_1}}(k_{\uprho}) & = & \left(\upkappa_{1}\text{sgn}\left(x-x_{0}\right)\hat{e}_{\uprho} - |k_{\uprho}|\hat{e}_{x}\right)\cdot\nonumber\\
&&\dfrac{k_{\uprho}|k_{\uprho}|}{\upkappa_{1}}\mathrm{H}_{~0}^{(1)}\left(k_{\uprho}\uprho\right)e^{i\upkappa_{1}|x-x_{0}|}~~,\label{eq:2}\\
\underline{\mathrm{f_2}}(k_{\uprho}) & = & \left(\upkappa_{1}\hat{e}_{\uprho} - |k_{\uprho}|\hat{e}_{x}\right)k_{\uprho}|k_{\uprho}|\cdot\nonumber\\
&&\dfrac{\upvarepsilon_{2}\upkappa_{1}-\upvarepsilon_{1}\upkappa_{2}}{\upkappa_{1}\left(\upvarepsilon_{2}\upkappa_{1}+\upvarepsilon_{1}\upkappa_{2}\right)}\mathrm{H}_{~0}^{(1)}\left(k_{\uprho}\uprho\right)e^{i\upkappa_{1}(x+x_{0})}~~,\vspace{2mm} \label{eq:3}\\
\IEEEeqnarraymulticol{3}{c}{
\upkappa_{1} = \sqrt{k_{01}^{2}-k_{\uprho}^{2}}~,~\upkappa_{2} = \sqrt{k_{02}^{2}-k_{\uprho}^{2}}~,}\label{eq:4}
\end{IEEEeqnarray}
with $\mathrm{H}_{~0}^{(1)}$ being the Hankel function of zero order and first kind and $k_{01}$, $k_{02}$, the wavenumbers of propagation in the air and lossy medium (ground) respectively.

Expressions (\ref{eq:1}) -- (\ref{eq:4})  expose the following difficulties, when coming to  the evaluation of the respective integrals through common Numerical Integration (NI) techniques:
\begin{itemize}
\item [-] The range of integration extends from $-\infty$ to $+\infty$, resulting in potential computational errors for large evaluation arguments.
\item [-] The Hankel function, $\mathrm{H}_{~0}^{(1)}$, exhibits a singularity at $k_{\uprho}=0$ and although it is proved that this is a logarithmic singularity\cite{Fikioris2013} and does not break the integral’s convergence{\footnote{since one can easily show: $\displaystyle \lim_{k_{\uprho} \to 0} k_{\uprho}\cdot \mathrm{H}_{~0}^{(1)}\left(k_{\uprho}\uprho\right)=0$\cite{Bourgiotis2015}.}}, it can affect the accuracy of the numerical integration results, when implemented in the computer.
\item [-] In addition, it is obvious that $k_{\uprho}=\pm k_{01}$ are also isolated singularities of (\ref{eq:2}), (\ref{eq:3}) and despite they are still integrable singularities\cite{Fikioris2013}{\footnote{it is a square root integrable singularity that applies to Rule 1 of\cite{Fikioris2013}.}}, a sufficient small range around those points must be excluded, when numerically evaluating (\ref{eq:1}) in the computer. As argued in\cite{Chrysostomou2016}, doing so may severely affect the accuracy of the results.
\end{itemize}

Of course, the above mentioned accuracy issues, are of practical importance, only as far as the \emph{Scattered Field}, $\underline{E}^{R}$, is concerned, for which no analytic formula exists. For the LOS field, a closed-form expression does exist as following:
\begin{IEEEeqnarray}{rCl}
\IEEEeqnarraymulticol{3}{c}{
\underline{E}^{LOS} = - \dfrac{i\upomega p}{4\uppi}\cdot e^{ik_{01}r}\times}\nonumber \\
&&\times\left\lbrace\left(\dfrac{-i\upomega \upmu_1}{2r_1} + \dfrac{3\upzeta}{2r_{1}^{2}} - \dfrac{3}{2i\upomega \upvarepsilon_{1} r_{1}^{3}}\right)\right. \sin 2\uptheta_{1}\cdot \hat{e}_\uprho +\label{eq:5}\\
&&\left.\left[\dfrac{i\upomega \upmu_{1}}{r_{1}}\sin^{2}\uptheta_{1} + \left(\dfrac{\upzeta}{r_{1}^{2}} -\hspace{-1mm}\dfrac{2}{i\upomega \upvarepsilon_{1} r_{1}^{3}} \right)\hspace{-1mm}\left(\cos 2\uptheta_{1} + \cos^{2}\uptheta_{1}\right) \right]\hspace{-1mm}\hat{e}_{x}\right\rbrace,\nonumber
\end{IEEEeqnarray}
\noindent which is found by solving the problem of an isolated hertzian dipole  source, in free space, located at point A of Fig. \ref{fig:1} and expressed in its local coordinate system $\left(\mathrm{O\equiv A}\right)$\cite{Fikioris1982}. However, for verification purposes, we will also briefly examine the integral representation for the LOS field as well. Note that (\ref{eq:5}) reflects the exact solution of the problem, encompassing both the near field and far field components and is expressed in the cylindrical coordinate system. Also, $\upzeta = \sqrt{\upmu_1/\upvarepsilon_1}$, the wave impedance of the medium above ground (free space or air in our case, i.e. $\upzeta\simeq 120\uppi~\Omega$).

\subsection{Reformulated Integral Expressions for the EM Field}\label{subsec:new_formulas}
We now focus on the scattered field, i.e. the second integral expression of (\ref{eq:1}), which may be written as
\begin{IEEEeqnarray}{C}
\underline{E}^{R} = -\dfrac{ip}{8\uppi\upvarepsilon_{0}\upvarepsilon_{1}}\left(I_{1}+\overbrace{I_{2}+I_{3}}^{I_{23}}\right),\label{eq:6}
\end{IEEEeqnarray}\vspace{-3mm}
\begin{subnumcases}{\hspace{-14mm}}
I_{1} = \int_{-k_{01}}^{+k_{01}}\underline{\mathrm{f_2}}(k_{\uprho})\mathrm{d}k_{\uprho},\label{eq:7a}\\
I_{2} = \int_{-\infty}^{-k_{01}}\underline{\mathrm{f_2}}(k_{\uprho})\mathrm{d}k_{\uprho},\label{eq:7b}\\
I_{3} = \int_{+k_{01}}^{+\infty}\underline{~\mathrm{f_2}}(k_{\uprho})\mathrm{d}k_{\uprho}.\label{eq:7c}
\end{subnumcases}

Starting with (\ref{eq:7a}), we perform a simple variable transform, $k_{\uprho}=k_{01}\sin\xi$, which apparently maps the $\left[-k_{01}, +k_{01}\right]$ range to $\left[-\uppi/2, +\uppi/2\right]$. With this transform, (\ref{eq:4}) is translated to
\begin{equation}
\upkappa_{1} = k_{01}\cos\xi,~~\upkappa_{2} = \sqrt{k_{02}^{2}-k_{01}^{2}\sin^{2}\xi}~.\label{eq:8}
\end{equation}
Ultimately and if we also take into consideration the definition for $\underline{\mathrm{f_2}}$, as given by (\ref{eq:3}), the expression for $I_{1}$ becomes

\begin{IEEEeqnarray}{l}
I_{1}\hspace{-1mm}=\hspace{-1mm}k_{01}^{3}\hspace{-1mm}\int_{-\frac{\uppi}{2}}^{+\frac{\uppi}{2}}\hspace{-2mm}
\begin{array}{l}
\left(\cos\xi ~\hat{e}_\uprho - \vert\sin\xi\vert ~\hat{e}_{x}\right)\cdot\sin\xi\vert\sin\xi\vert\cdot\\
R_{\lVert}\left(\xi\right)\cdot\mathrm{H}_{~0}^{(1)}\left(\uprho {k_{01}\sin\xi}\right)\cdot e^{ik_{01}(x+x_{0})\cos\xi}~\mathrm{d}\xi\IEEEeqnarraynumspace\\
\end{array}\label{eq:9}\\
\nonumber\\
\mathrm{with}\quad R_{\parallel}\left(\xi\right) = \frac{\upvarepsilon_{2} k_{01}\cos\xi - \upvarepsilon_{1}\sqrt{k_{02}^{2} - k_{01}^{2}\sin^{2}\xi}}{\upvarepsilon_{2} k_{01}\cos\xi + \upvarepsilon_{1}\sqrt{k_{02}^{2} - k_{01}^{2}\sin^{2}\xi}}~,\label{eq:10}
\end{IEEEeqnarray}

\noindent or equivalently it may be written as
\begin{IEEEeqnarray}{rcl}
I_{1}& = &k_{01}^{3}\left\lbrace\int_{0}^{\frac{\uppi}{2}}\left[
\begin{array}{l}\hspace{-1mm}
\left(\cos\xi ~\hat{e}_\uprho - \sin\xi ~\hat{e}_{x}\right)\cdot\sin^{2}\xi\cdot R_{\lVert}\left(\xi\right)\cdot\\
\mathrm{H}_{~0}^{(1)}\left(\uprho {k_{01}\sin\xi}\right)\cdot e^{ik_{01}(x+x_{0})\cos\xi}\\
\end{array}\hspace{-1mm}\right]\mathrm{d}\xi - \right.\nonumber\\
\nonumber\\
& - & \left.\int_{\hspace{-0.5mm}-\frac{\uppi}{2}}^{\hspace{0.5mm}0}\hspace{-1mm}\left[
\begin{array}{l}\hspace{-1mm}
\left(\cos\xi ~\hat{e}_\uprho + \sin\xi ~\hat{e}_{x}\right)\cdot\sin^{2}\xi\cdot R_{\lVert}\left(\xi\right)\cdot\\
\mathrm{H}_{~0}^{(1)}\left(\uprho {k_{01}\sin\xi}\right)\cdot e^{ik_{01}(x+x_{0})\cos\xi}\\
\end{array}\hspace{-1mm}\right]\mathrm{d}\xi\ \right\rbrace\ \hspace{-1mm}. \IEEEeqnarraynumspace \label{eq:11}
\end{IEEEeqnarray}
We may further elaborate on (\ref{eq:11}), if we make use of the following properties for the Hankel function\cite{Arfken2005, Olver1964}, namely,
\begin{IEEEeqnarray}{rCl}
\mathrm{H}_{~0}^{(1)}(z) + \mathrm{H}_{~0}^{(2)}(z) & = & \phantom{-}2\mathrm{J}_{0}(z),\label{eq:12}\\
\mathrm{H}_{~0}^{(1)}(ze^{i\uppi}) & = & - \mathrm{H}_{~0}^{(2)}(z),\label{eq:13}
\end{IEEEeqnarray}
(with the latter implying an analytic continuation of $\mathrm{H}^{(1)}_{~0}$in the upper half plane) and also observe from (\ref{eq:10}) that the reflection coefficient $R_{\parallel}\left(\xi\right)$ is an even function, with respect to $\xi$. Overall, we get
\begin{IEEEeqnarray}{l}
I_{1}\hspace{-0.5mm} =\hspace{-0.5mm} 2k_{01}^{3}\hspace{-1mm}\int_{0}^{\frac{\uppi}{2}}\hspace{-1mm}\left[\hspace{-1mm}
\begin{array}{l}
\left(\cos\xi ~\hat{e}_\uprho - \sin\xi ~\hat{e}_{x}\right)\cdot\sin^{2}\xi\cdot R_{\lVert}\left(\xi\right)\hspace{-0.5mm}\cdot\\
\mathrm{J}_{0}\left(\uprho {k_{01}\sin\xi}\right)\cdot e^{ik_{01}(x+x_{0})\cos\xi}\\
\end{array}\hspace{-1.5mm}\right]\hspace{-0.5mm}\mathrm{d}\xi,\IEEEeqnarraynumspace\label{eq:14}
\end{IEEEeqnarray}
where $\mathrm{J}_{0}$ denotes the zero order Bessel function.

For the integrals $I_{2}$ and $I_{3}$ we follow a similar approach. Particularly, in (\ref{eq:7b}) we apply the variable transform \mbox{$k_{\uprho}=k_{01}\cosh\xi$}, while in (\ref{eq:7c}) we set $k_{\uprho}=-k_{01}\cosh\xi$. In both cases, the original ranges of integration, $[-\infty, -k_{01}]$ and $[k_{01},+\infty]$, in the $k_{\uprho}$ domain, are mapped to $[0, +\infty]$ in the domain of $\xi$. Moreover, (\ref{eq:4}) becomes
\begin{equation}
\upkappa_{1} = ik_{01}\sinh\xi,~~\upkappa_{2} = \sqrt{k_{02}^{2}-k_{01}^{2}\cosh^{2}\xi}~.\label{eq:15}
\end{equation}
Performing the necessary calculations and also using (\ref{eq:12}), (\ref{eq:13}), we may combine the results for $I_{2}$ and $I_{3}$ as

\begin{align}
I_{23}\hspace{-1mm} =\hspace{-1mm}\frac{2k_{01}^{3}}{i}\hspace{-2mm}\int_{\hspace{0.2mm}0}^{\infty}\hspace{-1mm}\left[\hspace{-2mm}
\begin{array}{l}
\left(i\sinh\xi ~\hat{e}_\uprho - \cosh\xi ~\hat{e}_{x}\right)\cdot\cosh^{2}\xi\cdot R^{\prime}_{\lVert}\left(\xi\right)\hspace{-0.5mm}\cdot\\
\mathrm{J}_{0}\left(\uprho {k_{01}\cosh\xi}\right)\cdot e^{-k_{01}(x+x_{0})\cos\xi}\\
\end{array}\hspace{-2mm}\right]\hspace{-0.5mm}\mathrm{d}\xi,\label{eq:16}
\end{align}
where $I_{23} = I_{2} + I_{3}$ and the reflection coef. $R^{\prime}_{\parallel}$ is given by
\begin{equation}
R^{\prime}_{\parallel}\left(\xi\right) = \frac{i\upvarepsilon_{2} k_{01}\sinh\xi - \upvarepsilon_{1}\sqrt{k_{02}^{2} - k_{01}^{2}\cosh^{2}\xi}}{i\upvarepsilon_{2} k_{01}\sinh\xi + \upvarepsilon_{1}\sqrt{k_{02}^{2} - k_{01}^{2}\cosh^{2}\xi}}~.\label{eq:17}
\end{equation}
Substituting (\ref{eq:14}) and (\ref{eq:16}) to (\ref{eq:6}), we reach to an integral formula for the scattered field, $\underline{E}^{R}$, suitable for numerical calculations. With a similar process for the first integral of (\ref{eq:1}), we get the equivalent expression for the LOS field. Overall, the redefined integral expressions for the direct and scattered fields are given by
\begin{IEEEeqnarray}{l}
\underline{E}^{LOS} = \frac{-ipk_{01}^{3}}{4\uppi\upvarepsilon_{0}\upvarepsilon_{1}}~\cdot\nonumber\\
\cdot~\left\lbrace\int_{0}^{\frac{\uppi}{2}}\left[
\begin{array}{l}\hspace{-1mm}
\left(\mathrm{sgn}(x-x_{0})\cdot\cos\xi ~\hat{e}_\uprho - \sin\xi ~\hat{e}_{x}\right)\cdot\\
\sin^{2}\xi\cdot\mathrm{J}_{0}\left(\uprho {k_{01}\sin\xi}\right)\cdot e^{ik_{01}\vert x-x_{0}\vert\cos\xi} \\
\end{array}\hspace{-1mm}\right]\mathrm{d}\xi\right. - \nonumber\\
\nonumber\\
\hspace{-1mm}- i\hspace{-1mm}\left.\int_{0}^{^{\infty}}\hspace{-1.5mm}\left[
\begin{array}{l}\hspace{-1.5mm}
\left(i\mathrm{sgn}(x-x_{0})\cdot\sinh\xi ~\hat{e}_\uprho - \cosh\xi ~\hat{e}_{x}\right)\cdot\\
\cosh^{2}\xi\hspace{-0.5mm}\cdot\hspace{-0.5mm}\mathrm{J}_{0}\left(\uprho {k_{01}\cosh\xi}\right)\hspace{-0.5mm}\cdot\hspace{-0.5mm} e^{-k_{01}\vert x-x_{0}\vert\sinh\xi}\\
\end{array}\hspace{-1.5mm}\right]\hspace{-1mm}\mathrm{d}\xi\hspace{-0.5mm} \right\rbrace,\label{eq:18}\IEEEeqnarraynumspace
\end{IEEEeqnarray}

\begin{IEEEeqnarray}{l}
\underline{E}^{R} = \frac{-ipk_{01}^{3}}{4\uppi\upvarepsilon_{0}\upvarepsilon_{1}}~\cdot\nonumber\\
\cdot~\left\lbrace\int_{0}^{\frac{\uppi}{2}}\left[
\begin{array}{l}\hspace{-1mm}
\left(\cos\xi ~\hat{e}_\uprho - \sin\xi ~\hat{e}_{x}\right)\cdot\sin^{2}\xi\cdot R_{\lVert}\left(\xi\right)\cdot\\
\mathrm{J}_{0}\left(\uprho {k_{01}\sin\xi}\right)\cdot e^{ik_{01}(x+x_{0})\cos\xi} \\
\end{array}\hspace{-1mm}\right]\mathrm{d}\xi\right. - \nonumber\\
\nonumber\\
\hspace{-1mm}- i\left.\hspace{-0.5mm}\int_{0}^{^{\infty}}\hspace{-1mm}\left[
\begin{array}{l}\hspace{-1.5mm}
\left(i\sinh\xi ~\hat{e}_\uprho - \cosh\xi ~\hat{e}_{x}\right)\hspace{-0.5mm}\cdot\hspace{-0.5mm}\cosh^{2}\xi\hspace{-0.5mm}\cdot\hspace{-0.5mm} R^{\prime}_{\parallel}\left(\xi\right)\hspace{-0.5mm}\cdot\hspace{-0.5mm}\\
\mathrm{J}_{0}\left(\uprho {k_{01}\cosh\xi}\right)\cdot e^{-k_{01}(x+x_{0})\sinh\xi}\\
\end{array}\hspace{-1.5mm}\right]\hspace{-0.5mm}\mathrm{d}\xi \right\rbrace.\label{eq:19}\IEEEeqnarraynumspace
\end{IEEEeqnarray}

An inspection of (\ref{eq:18}) and (\ref{eq:19}) might yield some useful insights. Equation (\ref{eq:18}) expresses the direct field as an integral expression over the dummy variable $\xi$, which is an auxiliary, transformed variable of the spectral domain coordinate $k_{\uprho}$. As required by the problem's geometry, the field is cylindrically symmetrical (no $\upphi$ - component) and it is expressed as a complex summation of contributions, originating from the dipole's location, hence the dependence of the field on the horizontal distance, $\uprho$ and the relative height difference $x - x_{0}$. Moreover, it is easy to identify that the x - component of the field is symmetrical, while the $\uprho$ - component is antisymetrical, with respect to the dipole's vertical position $x_0$, which is in accordance to the conventional solution of the dipole's problem, as evident from an inspection of (\ref{eq:5})\footnote{$E_\mathrm{x}\left(x_0 - h\right) = E_x\left(x_0 + h\right),~E_\uprho\left(x_0 - h\right) = - E_\uprho\left(x_0 + h\right),~~ \forall ~\uprho, a$}. The expression for the scattered field has a similar form and can be considered as the integral generalization of Fresnel's theory, due to the existence of $R_{\parallel}\left(\xi\right)$ and $R^{\prime}_{\parallel}\left(\xi\right)$ in (\ref{eq:19}) that act as reflection coefficients, whose values depend on the ground characteristics. Also, the field depends on the cumulative distance $x + x_{0}$, as if the source is located at the image point $\mathrm{A}^\prime$ of Fig. \ref{fig:1}.

Equations (\ref{eq:18}), (\ref{eq:19}) remedy the accuracy issues, mentioned in section \ref{subsec:orig_formulas}, above:
\begin{itemize}
\item [-] They utilize the zero order Bessel function, $\mathrm{J}_{0}$, instead of $\mathrm{H}_{\phantom{(}0}^{(1)}$ used in (\ref{eq:1}), which is a smooth, finite special function with  no singularity, whatsoever.
\item [-] The singularities at points $k_{\uprho}=\pm k_{01}$ have also been removed. Hence, no need to exclude any range around them is required, when using any kind of numerical integration technique, in order to calculate (\ref{eq:18}) and (\ref{eq:19}). 
\item [-] The result is expressed as the sum of two integrals, one bound definite integral, in the range $[0, \uppi/2]$ and a second improper integral, for which the range of integration extends from 0 to $\infty$. However, due to the presence of $e^{-k_{01}(x+x_{0})\sinh\xi}$, the second integrand is a fast decaying function, practically making the integral a bound limits one that is fast converging and easy to evaluate in the computer.
\end{itemize}

The above findings are also visible in the simulations that follow.

\subsection{Simulations Results and Comparisons}
\label{subsec:comparisons1}
\begin{figure}[!t]
\centerline{\includegraphics[width=\columnwidth]{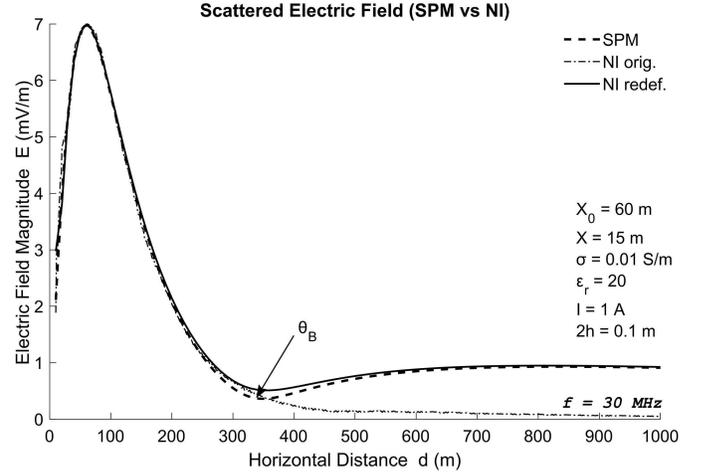}}
\caption{Comparison of Numerical Integration results for the received electric field
using: (i) redefined integral expressions, (\ref{eq:18}), (\ref{eq:19}) -- solid, (ii) original integral expressions (\ref{eq:1}) -- (\ref{eq:3}) (dash dotted). Scattered field values are depicted.}
\label{fig:2}
\end{figure}

The parameters for the various simulations (i.e. transmitter - receiver heights, ground parameters, operating frequency etc) are indicated within the figures and were selected such that a comparison with preceding, referenced results of\cite{Ioannidi2014, Bourgiotis2015, Chrysostomou2016} is possible. 

Fig. \ref{fig:2} exhibits the numerical evaluation (NI) for the scattered electric field, $\underline{E}^R$, using the redefined integral expression (\ref{eq:19}). It is compared to the equivalent values, obtained using the initial integral formulas for the electric field\footnote{expressed in the spectral domain}, introduced in\cite{CEMA2010, Ioannidi2014} that is by using (\ref{eq:1}) -- (\ref{eq:3}), above. Along with the NI results, the high frequency approximation values, are also superimposed. These values were obtained as in\cite{Bourgiotis2015, Chrysostomou2016}, i.e. through the application of the SPM method on (\ref{eq:1}). SPM is a useful asymptotic technique for the evaluation of complex integrals, particularly when the integrands expose rapidly changing phase components\footnote{For details regarding the conditions for applying the SPM, see\cite{Balanis1997}}.

As deduced in\cite{Chrysostomou2016}, SPM results are expected to be accurate in the far field, i.e. at least at distances over 10 -- 15 wavelengths, or above 100 -- 150m, for the 30MHz case, shown in Fig. \ref{fig:2}. Therefore, using the SPM data as the baseline, it is obvious that only the numerical evaluation of (\ref{eq:19}) achieves the required accuracy and this is noticeably evident for distances larger than the characteristic distance of the so-called Pseudo -- Brewster angle, defined as the angle of incidence, $\uptheta_\mathrm{B}$, where the reflected field is minimized\cite{Fikioris1982}\footnote{However this coincidence is not a general conclusion for every tested scenario; nevertheless it is a frequent case, which is why it's mentioned here.}. On the contrary, the numerical computation of (1) -- (3) fails to describe the electric field behavior, which may be attributed to the reasons analyzed in Section (\ref{subsec:orig_formulas}), above.

\begin{figure}[!t]
\centerline{\includegraphics[width=\columnwidth]{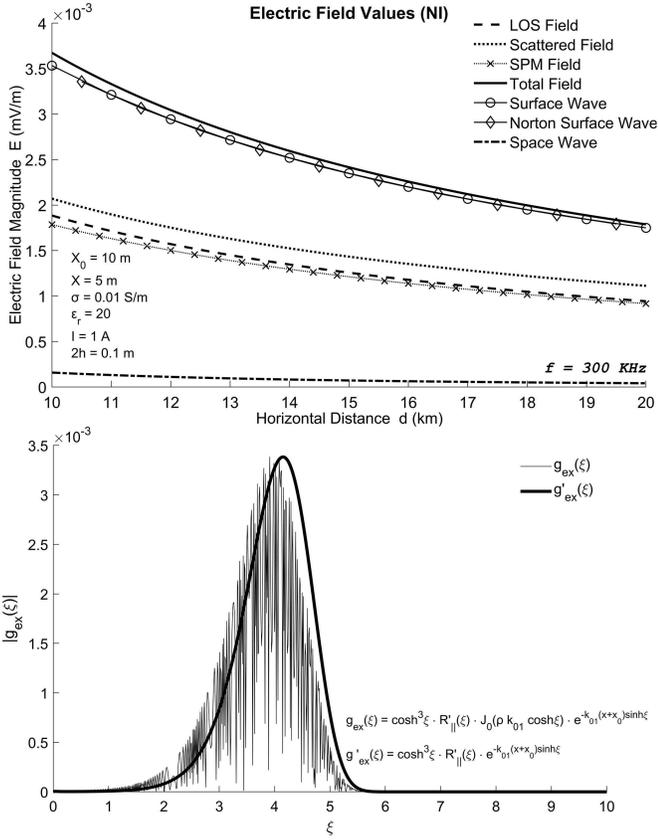}}
\caption{Numerical evaluation of the EM field at the LF/MF band; top: comparisons of various field types, bottom: integrand behavior of (\ref{eq:19}).}
\label{fig:3}
\end{figure}

In Fig. \ref{fig:3} we demonstrate various field types and components for the case of a hertzian dipole, radiating at 300 KHz, which is regarded as the frontier between the Low Frequency (LF) and Medium Frequency (MF) bands\cite{Fikioris1982}. For the LOS field we used (\ref{eq:5}), while the space wave was evaluated as in\cite{Sarkar2012}, i.e. by using the concept of the \emph{Fresnel Reflection Coefficient} for the reflected field. The scattered field\footnote{The terms scattered field and reflected field are not equivalent\cite{Sarkar2012}.} was numerically computed via (\ref{eq:19}).

Due to the small antenna heights and the long distances involved (10 -- 20 km), the space wave is expected to diminish\cite{Fikioris1982}. Therefore, the link is established primarily by means of the \emph{Surface Wave}, defined as the remaining field, after subtracting the geometrical optics field (or space wave) from the complete or \emph{Total Field}\cite{IEEEterms}. This is actually verified in the top plot of Fig. \ref{fig:3}, with the total field curve being very close to the surface wave results. As a confirmation of the validity of the results, our surface wave calculations are compared with Norton formulas\cite{Norton1936}. The respective curves are almost identical.

The bottom half of Fig. \ref{fig:3} displays the behavior of the integrand associated to the second integral of (\ref{eq:19}), i.e. the generalized integral over the $[0, \infty)$ range. Actually, we are dealing only with the x-component\footnote{This is the major field component for the considered problem\cite{Sarkar2012}.} of this integrand, denoted as function {{\verb+g+}}$_\mathrm{{ex}}(\xi)$ in Fig. \ref{fig:3} (the behavior for the $\uprho$-component is similar). The integrand is confined in a small window of the integration variable, $\xi$, outside of which and especially for large values of $\xi$, it is practically equal to zero. This is an outcome of the fact that the exponential function $e^{-k_{01}(x+x_{0})\sinh\xi}$ decreases much faster than the increase rate of $\cosh^3\xi$ \footnote{or alternatively $\lim_{\xi\to\infty}e^{-k_{01}(x+x_{0})\sinh\xi}\cdot\cosh^3\xi = 0$}. The bottom line is that the second integral of (\ref{eq:19}), essentially becomes a bound limits definite integral, which can be easily and accurately evaluated in the computer, using common numerical integration techniques. This gives our formulation a computational advantage.

In Fig. \ref{fig:3} it is also interesting to notice the fluctuating behavior of {{\verb+g+}}$_\mathrm{{ex}}(\xi)$. This is an outcome of the oscillating nature of the Bessel function $\mathrm{J}_{0}$. Its effect on {{\verb+g+}}$_\mathrm{{ex}}(\xi)$ is apparent by observing the bold line of Fig. \ref{fig:3} ({{\verb+g'+}}$_\mathrm{{ex}}(\xi)$ in the figure), which demonstrates how the integrand would behave, if it hadn't been for $\mathrm{J}_{0}$. Again, the confinement of the integrand within a ''narrow-band'' of the variable $\xi$ is apparent. It also seems that {{\verb+g'+}}$_\mathrm{{ex}}(\xi)$ acts like a slightly-shifted envelope function of {{\verb+g+}}$_\mathrm{{ex}}(\xi)$. However, notice that this is a normalized illustration of {{\verb+g'+}}$_\mathrm{{ex}}(\xi)$, to the respective magnitude of {{\verb+g+}}$_\mathrm{{ex}}(\xi)$, since the order of magnitude between the two is totally different.

\begin{figure}[!t]
\centerline{\includegraphics[width=\columnwidth, height=0.98\columnwidth]{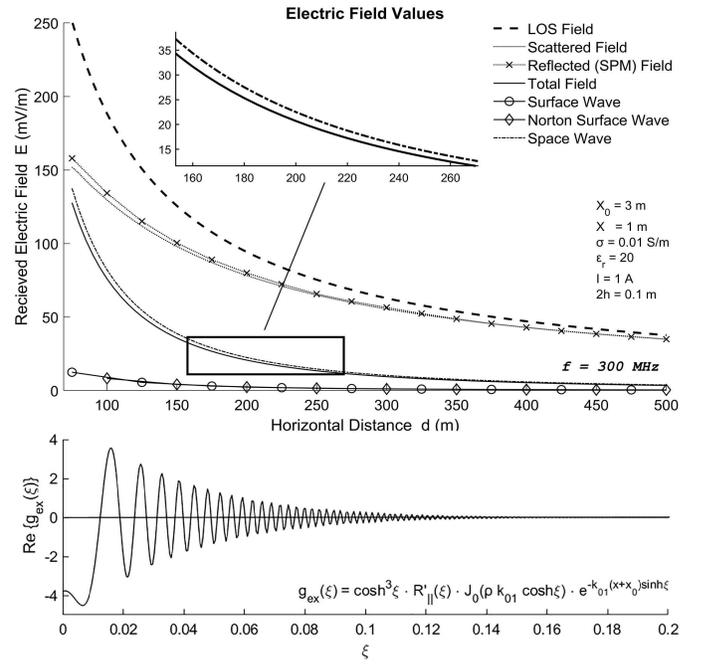}}
\caption{Numerical evaluation of the EM field at the VHF/UHF band; top: comparisons of various field types, bottom: integrand behavior of (\ref{eq:19}).}
\label{fig:4}
\end{figure}

The simulations of Fig. \ref{fig:3}	are now repeated for a high frequency scenario in the VHF / UHF band. The source and observation points are located even closer to the ground level, in an attempt to detect meaningful surface wave values, if possible, in this higher frequency case. Nevertheless, as illustrated in Fig. \ref{fig:4}, this is a situation where the space wave almost completely dictates the field behavior. The pursued surface wave becomes very quickly negligible and this is actually in accordance with Norton's predictions, where the large values for the so-called \emph{Numerical Distance}, results in very small values for the attenuation coefficient, hence small surface wave figures in the high frequency regime\cite{Fikioris1982}. These results are also a validation of the SPM method, which as mentioned in Section \ref{sec:introduction}, it emerges as the asymptotic solution for the complete problem, in the high frequency case.

\begin{figure}[!t]
\centerline{\includegraphics[width=\columnwidth]{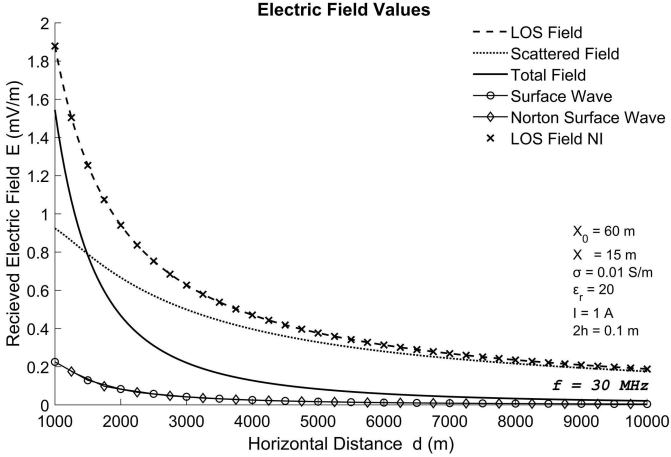}}
\caption{Electric field components at the frequency of 30 MHz.}
\label{fig:5}
\end{figure}

Finally, notice in the bottom graph of Fig. \ref{fig:4}, how quickly {{\verb+g+}}$_\mathrm{{ex}}(\xi)$ vanishes (in this case the real part is shown), making thus the convergence of (\ref{eq:19}) very fast. Moreover, due to the alternating positive and negative values, it is expected that the effect of {{\verb+g+}}$_\mathrm{{ex}}(\xi)$ on the overall result will be insignificant. The same arguments hold for the $\uprho$-component of (\ref{eq:19}), justifying the small observed values, as far as the surface wave field is concerned. Put it differently, for the case shown in Fig. \ref{fig:4}, the major contribution in (\ref{eq:19}) comes from a narrow area around the Stationary Point, which in this problem lies within the $\left[-\uppi/2, +\uppi/2\right]$ range\cite{Chrysostomou2016}. This contribution yields the reflected field in an asymptotic sense, as first shown in\cite{Ioannidi2014} with the application of the SPM method. In the rest of the integration range, the integrand is related with the surface wave\cite{IEEEterms} and exposes a behavior similar to Fig. \ref{fig:4}, thus having minimum impact to the final result. This was a major assumption for the application of the SPM method in\cite{Ioannidi2014}, which is now numerically validated in this high frequency scenario.

As a last validation, in Fig. \ref{fig:5} we demonstrate various field components for the exact scenario, illustrated in Fig. 4 of\cite{Ioannidi2014}. The simulation parameters are as those of Fig. \ref{fig:2}, except for the horizontal distance range. In\cite{Ioannidi2014} only the Norton's surface wave was evaluated, whereas here we also compare with the NI results. Moreover, we perform a comparison between the analytic expression for the LOS field and its equivalent integral form (''LOS field NI'' in Fig. \ref{fig:5}), as both given in Section \ref{sec:formulation} by (\ref{eq:5}) and (\ref{eq:18}) respectively. Again, our numerical evaluation for the surface wave is more or less identical with Norton's values. No needless to say that we also achieve a perfect match between (\ref{eq:5}) and (\ref{eq:18}), essentially meaning that our redefined integral formulation for the EM field, described in Section \ref{subsec:new_formulas}, is effective and accurate. In other words, the drawbacks, associated with the original integral expressions of Section \ref{subsec:orig_formulas}, do seem to have been mitigated.
\begin{figure}[!t]
\centerline{\includegraphics[width = \columnwidth]{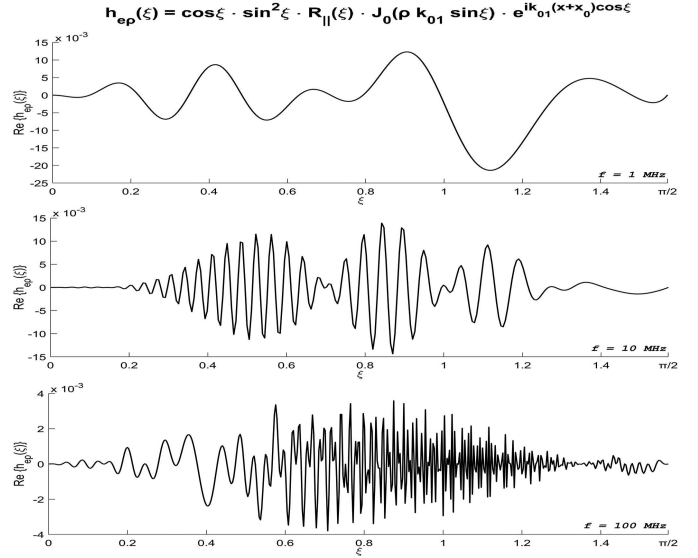}}
\caption{Integrand behavior of (\ref{eq:19}). The real part of the $\uprho$-component of the first integral expression of (\ref{eq:19}) is illustrated for $f = 1, 10, 100$ MHz. The horizontal distance $\uprho$ is 1 km and the T--R heights are $X_{0}=60$ m and $X = 15$ m, respectively.}
\label{fig:6}
\end{figure}
\begin{table}[!t]
\caption{Convergence Time}
\label{table:I}
\vspace{-2mm}
\centering
\setlength{\tabcolsep}{6pt}
{\renewcommand{\arraystretch}{1.4}
\begin{tabular}{|@{\hskip1pt}r@{\hskip2pt}|r@{.}l|r@{.}l|r@{.}l|r@{.}l|r@{.}l|r@{.}l|}
\multicolumn{1}{@{\hskip1pt}c@{\hskip2pt}}{•}&\multicolumn{12}{c}{Relative Tolerance}\\
\cline{2-13}
\multicolumn{1}{@{\hskip1pt}c@{\hskip2pt}}{\textit{f}}&\multicolumn{6}{|c}{Adaptive Simpson's}&\multicolumn{6}{|c|}{Trapezoidal}\\
\cline{2-13}
\multicolumn{1}{@{\hskip1pt}c@{\hskip2pt}}{(MHz)}&
\multicolumn{2}{|c}{$10^{-3}$}&
\multicolumn{2}{|c}{$10^{-6}$}&
\multicolumn{2}{|c|}{$10^{-9}$}&
\multicolumn{2}{|c}{$10^{-3}$}&
\multicolumn{2}{|c}{$10^{-6}$}&
\multicolumn{2}{|c|}{$10^{-9}$}\\
\hline
1 & 3&88 &6&02 & 18&73 & 3&43 &43&14 & 969&11\\
\hline
3 & 4&48 &6&98 & 20&84 & 4&62 &50&94 & 1272&52\\
\hline
10 & 5&11 &7&92 & 21&69 & 7&43 &69&27 & 1921&38\\
\hline
30 & 6&82 &9&02 & 25&16 & 14&38 &114&05 & 2923&07\\
\hline
80 & 9&80 &14&93 & 32&15 & 31&25 &240&26 & 6748&43\\
\hline
100 & 11&00 &16&20 & 39&55 & 39&46 &354&70 & 9560&96\\
\hline
300 & 21&29 &35&77 & 61&24 & 57&60 &520&40 & 15437&25\\
\hline
1000 & \phantom{0}58&44 &103&55 & 156&89 & 126&47 &973&65 & 32240&70\\
\hline
\multicolumn{13}{p{225pt}}{Convergence times in milliseconds (ms). The horizontal distance was set to $\uprho = 1$ km. The rest of the problem parameters: $X$, $X_{0}$, $\upsigma$, $\upvarepsilon_{r}$, I, 2h, were set as in Fig. \ref{fig:5}. Simulations performed on a 64bit, Quad Core CPU@2.60 GHz, 16.0 GB RAM platform, using MatLab.}\\
\end{tabular}}
\label{tab1}
\end{table}

We close this section with a few comments regarding the method's efficiency. The convergence time of the method depends on four key aspects; a) the utilized HW and SW platform, b) the selected NI algorithm for the calculation of (\ref{eq:18}), (\ref{eq:19}), c) the required error tolerance and d) the problem parameters; especially the frequency of operation, for a given Transmitter - Receiver (T--R) distance and altitudes, or on the electric distance $k_{01}r$, when their combined effect is accommodated. The first three factors seem quite reasonable. Regarding the fourth one, which in first glance may seem less relevant, Fig. \ref{fig:6} provides a good reasoning. It displays the behavior of the first integrand of (\ref{eq:19})\footnote{the real part of the $\uprho$-component, just as in Fig. \ref{fig:3} we expose the magnitude of the x-component of the second integrand of (\ref{eq:19}).}, shown as {{\verb+h+}}$_\mathrm{{e\uprho}}(\xi)$ in the figure. It is obvious that higher frequencies contribute to additional oscillations and this is not surprising if one observes that {{\verb+h+}}$_\mathrm{{e\uprho}}(\xi)$ includes a Bessel and a phase function (cosine or sine function when the real or imaginary part is considered), which are increasingly fluctuating for larger arguments, as occurs in this case, when the frequency $f = \frac{k_{01}\cdot c}{2\pi}$ increases. Hence, one might expect that more steps or intervals are required, for the NI algorithm to achieve a given error threshold.

Table \ref{tab1} demonstrates the measured performance of our method, at various frequencies, utilizing two widely used NI techniques for the evaluation of (\ref{eq:19}), namely the Adaptive Simpson's and the Trapezoidal method\cite{Thomas2010}. We are able to calculate the fields at almost an arbitrary accuracy level  and at very reasonable computational times.\footnote{$10^{-12}$ or even lower error is also achievable at the expense of computational time.} Table \ref{tab1}, also exposes the effectiveness of adaptive quadrature NI techniques for the evaluation of such ill-behaved, rapidly fluctuating functions, such as {\verb+g+}$_\mathrm{{ex}}(\xi)$ and {\verb+h+}$_\mathrm{{e\uprho}}(\xi)$ of Figs. \ref{fig:3}, \ref{fig:4} and \ref{fig:6}\cite{McKeeman1962, Hillstrom1970}. Finally, the effect of the frequency on the convergence times is apparent. Depending on the required error allowance, it seems that above a certain frequency level, the selection of an adaptive quadrature technique, like the Adaptive Simpson's in our case, might be necessary for getting timely results. 

\section{Evaluating a Novel Asymptotic Solution to the Sommerfeld's Problem}
\label{sec:asymptotic_eval}
Now that we have a solid method for the numerical calculation of Sommerfeld Integrals, we may use it to examine a newly introduced asymptotic solution to the well--known Sommerfeld Radiation Problem. The method was first presented in\cite{Sautbekov2018} and briefly discussed below for ease of reference.

\subsection{Outline of the Asymptotic Method}
\label{subsec:synopsis}
Using the rigorous mathematical analysis of\cite{Sautbekov2018}, the field scattered by a planar interface, can be expressed as
\begin{IEEEeqnarray}{rCl}
\underline{E}^{R} & = & -\hat{e}_{\uptheta_{2}}\dfrac{pk_{01}^{3}}{2\upvarepsilon_{0}\upvarepsilon_{1}}\sqrt{\dfrac{-2i}{\uppi k_{01}\uprho}}\cdot e^{ik_{01}r_{2}\cos\upzeta_{p}}\cdot\nonumber\\
&&\cdot\sin^{^{\frac{3}{2}}}\hspace{-1mm}\uptheta_{2}\sin\frac{\upzeta_{p}}{2}R_{\parallel}\left(\uptheta_{2}\right)X\left(k_{01}r_{2}, -\upzeta_{p}\right),\label{eq:20}
\end{IEEEeqnarray}
where, with respect to Fig. \ref{fig:1}, $\hat{e}_{\uptheta_{2}}=\hat{e}_{\uprho}\cos\uptheta_{2} - \hat{e}_{x}\sin\uptheta_{2}$ refers to the unit vector, along the $\uptheta_2$ -- direction of a spherical coordinate system, whose origin is the dipole's image ($\mathrm{A}^\prime$) and $R_{\parallel}\left(\uptheta_{2}\right)$ is given by (\ref{eq:10}), for $\xi = \uptheta_{2}$. Moreover, in (\ref{eq:20}) $\upzeta_{p} = \xi_{p} - \uptheta_{2}$, where $\xi_{p}$ is the pole of $R_{\parallel}\left(\uptheta_{2}\right)$. Also, notice that (\ref{eq:20}) is derived under the usual case scenario, where $\upsigma\gg\upomega\upvarepsilon_{0}$, in which case $\xi_{p}$ may be approximated by
\begin{IEEEeqnarray}{rCl}
\xi_{\text{p}}& \simeq & \frac{\uppi}{2}+\sqrt{\frac{\upomega\upvarepsilon_{0}\upvarepsilon_{1}}{2\upsigma}}~\cdot\nonumber\\
&&\left\lbrace 1+\frac{\upomega\upvarepsilon_{0}\left(\upvarepsilon_{1}+\upvarepsilon_{2}\right)}{2\upsigma}-i\left[1-\frac{\upomega\upvarepsilon_{0}\left(\upvarepsilon_{1}+\upvarepsilon_{2}\right)}{2\upsigma}\right]\right\rbrace.\IEEEeqnarraynumspace\label{eq:21}
\end{IEEEeqnarray}

\begin{figure}[!t]
\centerline{\includegraphics[width = \columnwidth]{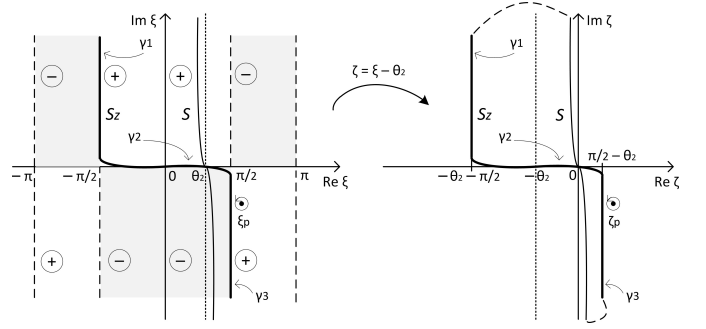}}
\caption{The contour of integration: a)$S_{z}$: original contour for $\underline{E}^{R}$ in the complex $\xi$-plane (left plot) and its $\upzeta$-plane mapping (right plot)  b)$S$: ``Etalon integral'' contour in the $\upzeta$-plane (right plot) and its $\xi$-plane mapping (left plot), c)$\xi_{p}$: relative position of the pole in the $\xi$-plane, d) $\upzeta_{p}$: relative position of the pole in the $\upzeta$-plane\cite{Sautbekov2018}.}
\label{fig:7}
\end{figure}

The most interesting part in (\ref{eq:20}) is special function $X$, the so-called 'Etalon Integral'\cite{Fock1945, Weinstein1969, Fock1946, Leontovic1946, Pedlosky2003, Sautbekov2010}. For parameters $k, \alpha$, it is defined as the contour integral
\begin{IEEEeqnarray}{rCl}
X\left(k,\alpha\right)\hspace{-1mm} & = & \hspace{-1mm}\frac{1}{4\uppi i}\int_{S}\hspace{-1mm}\frac{e^{ik\left(\cos\upzeta-\cos\alpha\right)}}{\sin\frac{\upzeta+\alpha}{2}}d\upzeta = \frac{e^{-i\frac{\uppi}{4}}}{\sqrt{2\uppi}}\int_{\infty\sin\frac{\alpha}{2}}^{2\sqrt{k}\sin\frac{\alpha}{2}}\hspace{-1mm}\large{e^{\frac{it^{2}}{2}}dt}\nonumber\\
& = & -\frac{1}{2}\text{sgn}\left(\text{Re}\left\lbrace\alpha\right\rbrace\right)+\frac{1}{2}\text{erf}\left(\sqrt{-2ik}\sin\frac{\alpha}{2}\right)\label{eq:22},
\end{IEEEeqnarray}
along path $S$ of  Fig. \ref{fig:7}\footnote{Regarding the notation in (\ref{eq:22}): $\infty\sin\frac{\alpha}{2} = \begin{cases}
+\infty, ~\sin\frac{\alpha}{2} > 0 \\
-\infty, ~\sin\frac{\alpha}{2} < 0
\end{cases}$}. The 'Etalon Integral' has useful properties and as shown in (\ref{eq:22}), it can be expressed in terms of Fresnel Integrals, which enable its easy evaluation via the well-known error function.

Keep in mind that to reach (\ref{eq:20}) the Saddle Point Method was used, in order to deform the original Sommerfeld contour of integration, $S_{z}$, into $S$, such that the expression for the Etalon Integral, (\ref{eq:22}), could be used. Therefore, the method is still another high frequency asymptotic method. The procedure is described in detail in\cite{Sautbekov2018}, of which Fig. 3 is replicated as Fig. \ref{fig:7}, in this manuscript, such that the relevant contours and the mapping process are briefly clarified. Finally, notice that at first glance, the pole $\xi_{p}$ seems not to influence the result, since it is kept outside the closed contour of Fig. \ref{fig:7} (right plot) and hence its residue is not considered. This is why the condition $\upsigma\gg\upomega\upvarepsilon_{0}$ that ensures the above argument, is important. However, the pole's relative location does have its meaning, as further described below.

It is also possible to further elaborate on (\ref{eq:22}), if one applies the large and small argument approximations for $\mathrm{erf}(z)$\cite{DLMF}. This results in
\begin{IEEEeqnarray}{rClr}
X\left(k,\alpha\right) & \simeq & -\sqrt{\frac{i}{2\uppi}}\frac{e^{\left[ik\left(1-\cos\alpha\right)\right]}}{2\sqrt{k}\sin\frac{\alpha}{2}}~, & ~~\sqrt{2k}|\sin\frac{\alpha}{2}|\gg 1 \label{eq:23},\IEEEeqnarraynumspace\\
X\left(k,\alpha\right) & \simeq & -\frac{1}{2}\text{sgn}\left(\alpha\right)+\sqrt{\frac{k}{2\uppi i}}\alpha~, & \frac{k\alpha^2}{2} < 1, \label{eq:24}\IEEEeqnarraynumspace
\end{IEEEeqnarray}
which when applied to (\ref{eq:20}), i.e. for $k = k_{01}r_{2}$ and $\alpha = -\upzeta_{p}$, they yield the following analytic expressions,

\begin{flalign}
\underline{E}^{R}\simeq\hspace{-1mm}
 -\hat{e}_{\uptheta_{2}} R_{\parallel}\left(\uptheta_{2}\right)\frac{pk_{01}^{2}}{4\uppi\upvarepsilon_{0}\upvarepsilon_{1}r_{2}}\sin\uptheta_{2}\cdot e^{ik_{01}r_{2}}~,\nonumber&&
\end{flalign}
\vspace*{-4mm}
\begin{flalign}
&&\sqrt{2k_{01}r_{2}}\cdot\sin\frac{\upvarphi}{2}\gg 1\hspace{2mm}\label{eq:25}
\end{flalign}
\begin{flalign}
\underline{E}^{R}\simeq\hat{e}_{x}\updelta\frac{pk_{01}^{3}}{4\upvarepsilon_{0}\upvarepsilon_{1}}\cdot\frac{1}{\sqrt{\uppi k_{01}\uprho}}e^{-\updelta k_{01}\left(x+x_{0}\right)}\cdot e^{i\left(k_{01}\uprho+\uppi/2\right)}~\cdot\nonumber&&
\end{flalign}
\vspace*{-4mm}
\begin{flalign}
\qquad~&\cdot\left[1+2i\sqrt{\frac{k_{01}\uprho}{\pi}}\cdot\updelta\left(1+k_{01}\uprho\updelta^{2}\right)\right],&\hspace{-4mm}
\begin{array}{r}
\upvarphi \rightarrow 0\\
k_{01}\uprho\updelta^{2} < 1
\end{array}
\label{eq:26}
\end{flalign}with $\updelta = \sqrt{\frac{\upomega\upvarepsilon_{0}\upvarepsilon_{1}}{2\upsigma}}\ll 1$ and $\upvarphi$ is the grazing angle of Fig. \ref{fig:1}. Notice that the second required condition of (\ref{eq:26}) refers to a quantity, which is essentially a kind of \emph{Numerical Distance} \cite{Norton1936}; more on it in Section \ref{subsec:comparisons2}, below.

Expression (\ref{eq:25}) indicates the geometric optics reflected field, emanating from  $\mathrm{A}^{\prime}$, the dipole's image point (Fig. \ref{fig:1}). It should be accurate for a long electric distance, $k_{01}{\cdot}r_{2}$, i.e. at the far field region, provided that at the same time the grazing angle $\upvarphi = \uppi/2 - \uptheta_{2}$ is not very small. In\cite{Ioannidi2014}, we also reached (\ref{eq:25}), using the SPM method. However, as stated in\cite{Chrysostomou2016}, the SPM required only the fulfillment of a large electric distance. The effect of the grazing angle was essentially overlooked and hence the propagation mechanism for the case of low height transmission link (where the angle of incidence is small) could not be highlighted. Pay attention to the fact that if (\ref{eq:25}) was absolutely accurate, even for sliding angles of incidence, just because of a high frequency transmitting source, the field to be received would essentially be imperceptible, since, in this case, the reflection coefficient, $R_{\parallel}$, approaches to $-1$ and $\underline{E}^{R}$ would simply cancel $\underline{E}^{LOS}$.

Regarding (\ref{eq:26}), we are given with an expression that describes the behavior of the scattered field, for sliding angles of incidence. Due to the existence of the exponentially decaying function, $e^{-\updelta k_{01}\left(x+x_{0}\right)}$, it is confined near the interface, as if it is a kind of a surface wave. The field is also spatially limited by the the required conditions, $\upvarphi \simeq 0, ~k_{01}\uprho\updelta^{2} < 1$, with the implications described in Section \ref{subsec:comparisons2}, below. However, pay attention to the fact that this is not a true surface wave, at least when one of the accepted definitions for a type of surface wave is considered\cite{IEEEterms}. It simply resembles a surface wave and this is a consequence of the boundary conditions and the pole's proximity to the contour of integration, as illustrated in Fig. \ref{fig:7}. We choose to call this a \emph{pseudo-Surface Wave}.


In the simulations that follow, we compare the closed-form asymptotic solution of\cite{Sautbekov2018}, i.e. (\ref{eq:20}), against the SPM-based solution of\cite{Ioannidi2014}, which essentially leads to the geometrical optics field expressions in the high frequency regime. The reference for our comparisons are the numerical integration results that we obtain for the EM field, using the methodology of Section \ref{sec:formulation}, above that is the evaluation of (\ref{eq:18}), (\ref{eq:19}) and the respective formulas for the magnetic field. In addition, we also examine and comment on the accuracy of the analytic expressions (\ref{eq:25}), (\ref{eq:26}).

\subsection{Simulation Results}
\label{subsec:comparisons2}
\begin{figure*}[!t]
\centerline{\includegraphics[width=\textwidth]{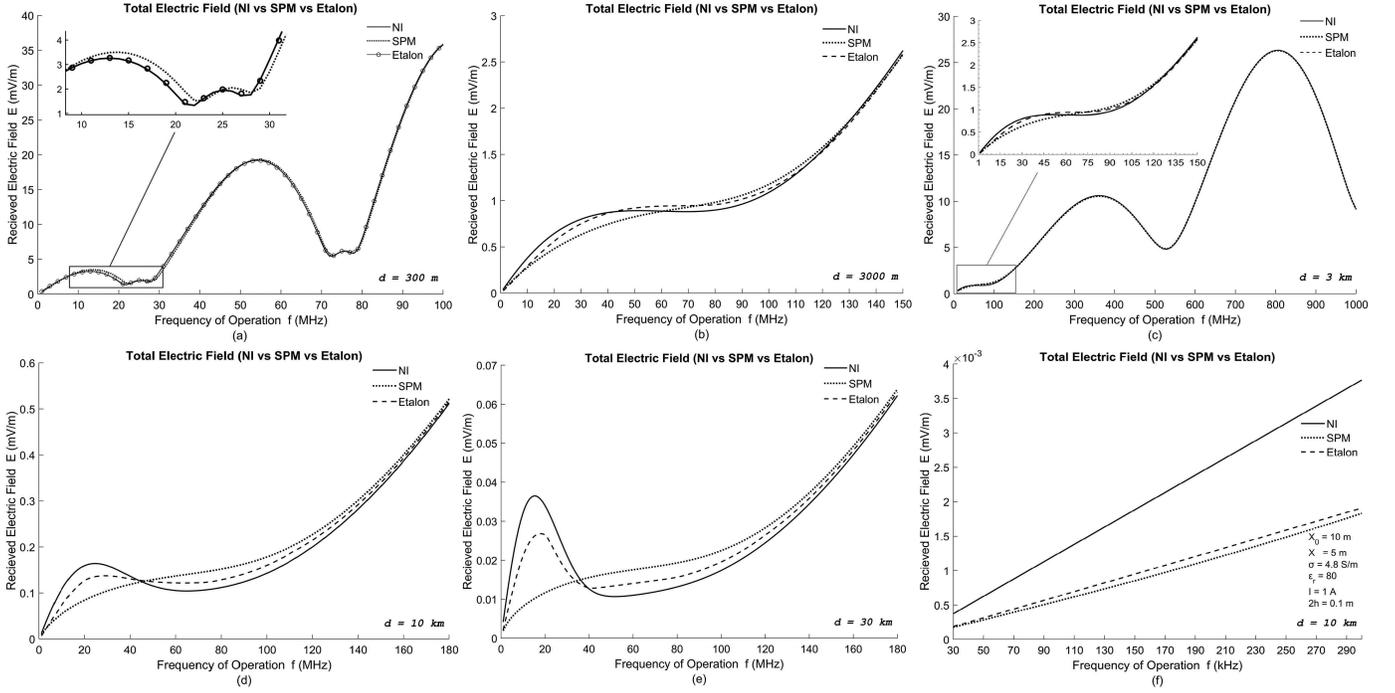}}
\caption{The variation of the total received electric field (magnitude) with respect to frequency ({\textbf{\it{f}}}), over various horizontal distance ({\textbf{\it{d}}}) scenarios, as predicted by: a) Numerical Integration (NI) of (\ref{eq:18}), (\ref{eq:19}) -- ''NI'', b) SPM - based asymptotic solution [18] -- ''SPM'' and c) asymptotic solution of [25] -- ''Etalon''.}
\label{fig:8}
\end{figure*}

\begin{figure*}[!t]
\centerline{\includegraphics[width=\textwidth]{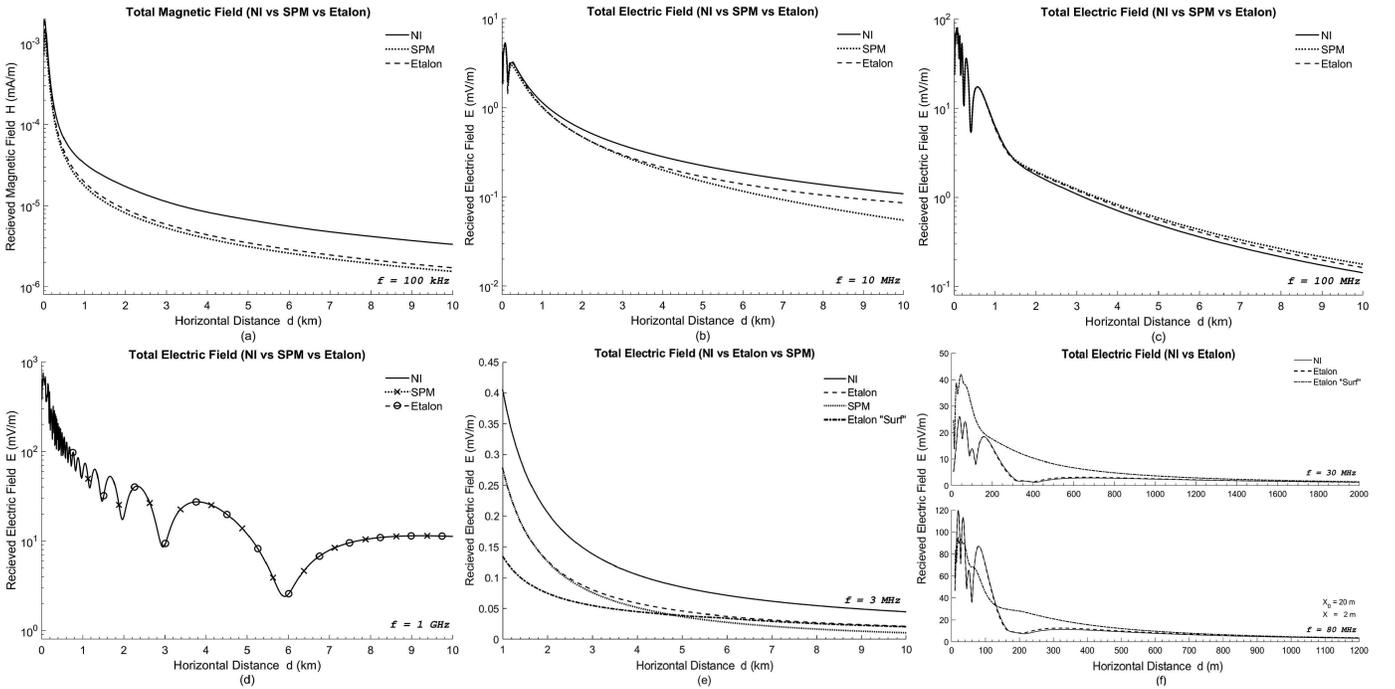}}
\caption{EM field magnitude over horizontal distance ({\textbf{\it{d}}}), for various distinct frequencies ({\textbf{\it{f}}}). The term Etalon ''Surf'', refers to the evaluation of (\ref{eq:26}). Fig. \ref{fig:9}(a), exhibits the magnetic field. The rest of the labeling convention of Fig. \ref{fig:8} applies.} 
\label{fig:9}
\end{figure*}

We exhibit two sets of simulations, in Figs. \ref{fig:8} and \ref{fig:9}, below. Fig. \ref{fig:8} demonstrates the effect of the frequency on the total received electric field, for a number of scenarios, regarding the Transmitter -- Receiver (T--R) horizontal distance, denoted with ''d'' in the respective plots. With the exception of Fig. \ref{fig:8}(f), the basic simulation parameters are shown in Table \ref{tab2}. The ground parameters, $\upvarepsilon_{r}$, $\mu $, $\upsigma $, are indicative for the case of sea water and do fulfill the basic requirement, $\upsigma\gg\upomega\upvarepsilon_{0}$, mentioned in Section \ref{subsec:synopsis}. The altitudes $X_{0}$ and $X$ are kept constant, at 60 m and 15 m respectively, however by increasing the horizontal distance, d (up to 30 km in Fig. \ref{fig:8}(e)), we may simulate sliding angles of incidence as well. Only in the case of Fig. \ref{fig:8}(f), where the frequencies involved are significantly lower, did we further lower the antennas' heights and this was done to examine the degree to which the methods are able to detect the so-called surface wave, which in this case should be more significant\cite{Fikioris1982}. We also focus on the far-field behavior. Finally, for the evaluation of the error function in (\ref{eq:22}), which due to (\ref{eq:20}) now includes a complex argument ($-\upzeta_{p}$), the algorithms described in\cite{Zaghloul2012, Zaghloul2017} were appropriately utilized, which very accurately evaluate such special functions in the complex plane.
\begin{table}
\centering
\caption{Simulations Parameters}
\label{table:II}
\setlength{\tabcolsep}{3pt}
\begin{tabular}{|p{25pt}|p{120pt}|p{65pt}|}
\hline
Symbol& 
Description& 
Value
\\
\hline
$f_{min} $& 
minimum frequency& 
1 MHz \\
$f_{max}$& 
maximum frequency & 
1 GHz \\
$X_{0}$& 
height of transmitting dipole& 
60 m \\
$X$& 
height of receiver's position& 
15 m \\
$I$& 
dipole's ipole current source& 
1 A \\
$2h$& 
dipole's length& 
0.1 m$^{\mathrm{a}}$ \\
$\upsigma $&
ground conductivity& 
4.8 S/m$^{\mathrm{b}}$ \\
$\upvarepsilon_{r}$& 
ground relative permitivity& 
80$^{\mathrm{b}}$ \\
$\mu $& 
ground permeability& 
$4\pi \times  10^{-7}$ H/m \\
& 
numerical integration technique& 
Adaptive Simpsons \\
& 
relative error tolerance& 
$10^{-6}$ \\
\hline
\multicolumn{3}{p{220pt}}{$^{\mathrm{a}}$much smaller than the wavelength $\lambda = c / f$}\\
\multicolumn{3}{p{220pt}}{$^{\mathrm{b}}$pertains for the case of see water}
\end{tabular}
\label{tab2}
\end{table}

The case of Fig. \ref{fig:8}(a) is indicative of non-near-ground-level communication. The T--R relative position is such that the angle of incidence is $\upvarphi\simeq 15^{\circ}$. It is evident that there is an almost perfect match between the results obtained numerically, labeled as ''NI'' in the plots and what is predicted by the newly introduced asymptotic solution (\ref{eq:20}), depicted via the ''Etalon'' indicated lines in Figs. \ref{fig:8} and \ref{fig:9}. It is also equally interesting that the older asymptotic SPM method, also yields similar results, which for frequencies around 20 MHz and above are consistently almost identical with what is numerically computed. Keep in mind that the SPM solution is essentially the expression given by (\ref{eq:25}), which in turn is derived as a special case of (\ref{eq:20}). However, the restriction for (\ref{eq:25}), namely $\sqrt{k_{01}r_{2}}\cdot\sin\frac{\upvarphi}{2}\gg 1$, is not strictly fulfilled in our case. For the scenario of Fig. \ref{fig:8}(a) it goes from 0.31@1MHz to 3.1@100MHz. At 20MHz it is about 1.4. Thus, it seems that (\ref{eq:25}) is an accurate analytic expression, for non--sliding angle of incidence reception, whose validity could be practically extended beyond the strict restrictions imposed for its derivation.

In Fig. \ref{fig:8}(b) the T--R distance is increased to 3 km and as a result the angle of incidence is radically reduced to $\upvarphi\simeq 1.43^{\circ}$. In this scenario, we do observe a discrepancy between the two asymptotic solutions and of both of them with the reference numerical integration (NI) results for (\ref{eq:19}). Of course, this discrepancy appears to be relatively small and if examined in a broader frequency range, as in Fig.\ref{fig:8}(c), it may be practically regarded negligible. Nevertheless, it is important to note the tendency of (20) to better follow (\ref{eq:19}), something that is even more apparent in the diagrams (d) and (e) of Fig.\ref{fig:8}. In these cases, the T--R distance is further increased to 10 km and 30 km, with the incidence angles now being as sliding as $\upvarphi\simeq 0.43^{\circ}$ and $\upvarphi\simeq 0.14^{\circ}$, respectively. Overall, compared with the solution of\cite{Ioannidi2014} (SPM based solution), the recently introduced asymptotic method in\cite{Sautbekov2018} (Etalon based solution) is a better estimate to the total solution of the Sommerfeld's radiation problem. It is also apparent that both methods smoothly converge to (\ref{eq:19}), in the high frequency regime, but the solution of\cite{Sautbekov2018} converges faster. On the contrary, Fig. \ref{fig:8}(f) verifies a somehow expected behavior. In lower frequencies, both methods fail to describe the propagation mechanism, for being unable to capture the effect of the surface wave, which in this scenario should be rather significant\footnote{Consider also the lower T--R heights, selected particularly in this case, such that the presence of the surface wave is further exaggerated.}. Indeed, (\ref{eq:20}) behaves only marginally better than the respective asymptotic formula of\cite{Ioannidi2014}, which as already stated, essentially yields the space wave component and ignores the contribution of the surface wave. The results of Fig. \ref{fig:8}(f) were somehow expected, since both solutions are based on the application of high freq. assymptotic methods (SPM vs Saddle Point) and may therefore yield accurate results only in the high frequency regime\cite{Bender1999}.

To confirm and further solidify the above arguments, in Fig. \ref{fig:9} we exhibit the field behavior from the perspective of a varying T--R distance. Starting from the low frequencies, in Fig. \ref{fig:9}(a)\footnote{For completeness, magnetic field values are given just for this case.} it is apparent that both asymptotic methods fail to produce accurate results. Actually, according to our detailed simulations, this situation holds true almost up to approximately 1 MHz. Moving, towards the HF frequency zone, Fig. \ref{fig:9}(b), the advantages of the newly introduced asymptotic solution show up. The difference between (\ref{eq:20}) and the previous SPM--based solution of\cite{Ioannidi2014} is more evident for large distances, where the effect of the scattered field is more significant; hence, the improvement that the ''Etalon'' function, $X$, yields in (\ref{eq:20}) becomes visible. If we further proceed to the VHF zone of Fig. \ref{fig:9}(c), we realize that both methods begin to converge and ultimately they coincide with the complete solution at even larger frequencies, as indicatively shown in Fig. \ref{fig:9}(d). At those frequencies and in accordance with what is known in the literature, the surface wave is almost negligible. Therefore, there is almost nothing extra left for special function $X$ to expose and (\ref{eq:20}) simply yields the reflected field, exactly as the  asymptotic solution of\cite{Ioannidi2014} does.

The last two diagrams of Fig. \ref{fig:9} are devoted to the investigation of (\ref{eq:26}), an interesting expression, which as mentioned in Section \ref{subsec:synopsis}, attributes surface wave characteristics to the near-ground-level scattered field. For this, extended simulations were run, whose outcomes are summarized in Figs. \ref{fig:9}(e), (f). Essentially, (\ref{eq:26}) does converge to (\ref{eq:20}), from which it was derived when $\upvarphi \to 0$. In the case of Fig. \ref{fig:9}(e), the convergence occurs approximately at 6 km, which for the selected T--R altitudes (Table \ref{table:II}), is equivalent to a grazing angle $\upvarphi$ of less than $1^\circ$. Of course, as already mentioned, in this frequency band ($\le$3 MHz), (\ref{eq:20}) and therefore (\ref{eq:26}) as well, are not totally accurate approximations of the complete solution. They are simply better estimates compared to the SPM and hence the visible gap between them and the NI results in Fig. \ref{fig:9}(e).

The same scenario is repeated at 30 MHz, illustrated in the top diagram of Fig. \ref{fig:9}(f). This time the convergence is achieved before 1400 m, where the grazing angle $\upvarphi$ is just above $3^{\circ}$. In general, for the same T--R heights, we observe a tendency of slightly increasing grazing angles, as the frequency increases. However, in this case, the numerical distance, $k_{01}\uprho\updelta^{2}$, quickly approaches the value of 1, due to its dependence on the square of the frequency. Compared to the 3 MHz case and for a certain horizontal distance $\upvarrho$, it is now 100 times larger, essentially shrinking the range for which (\ref{eq:26}) is accurate. A similar behavior is observed in the bottom diagram of Fig. \ref{fig:9}(f). Due to the lower antenna heights, we have a match at about 650 m, where $\upvarphi\simeq 2^{\circ}$. Nevertheless, because of the higher frequency, the validity of (\ref{eq:26}) is now limited to an even shorter distance range. We thus reach to the conclusion that this pseudo-surface wave has local significance and its existence is highly dependent on the value of the numerical distance, $k_{01}\uprho\updelta^{2}$. When the numerical distance exceeds 1, it essentially disappears.


\section{Conclusion and Future Research}
\label{sec:conclusion}
We demonstrated an efficient method for the numerical evaluation of Sommerfeld Integrals in the spectral domain. The method proves fast and accurate and when applied to the evaluation of the EM field, of a radiating vertical dipole above flat lossy ground, it fits very well with existing asymptotic solutions and Norton's results.

With a reference numerical method, to accurately evaluate the integral representation to the Sommerfeld's ratiation problem, i.e. (\ref{eq:18}), (\ref{eq:19}), we then focused on the evaluation of a recently developed asymptotic solution\cite{Sautbekov2018}. The solution uses the saddle point method and utilizes the properties of the so--called 'Etalon Integral', as a means to increase the accuracy of the results. Through extensive simulations, we verified that for the usual case, where $\upsigma\gg\upomega\upvarepsilon_{0}$, the method does succeed to provide better estimates, as compared with a more basic asymptotic approach, which is based on the application of the stationary phase method. Moreover, further asymptotic properties for the 'Etalon Integral' allowed us to reach analytic formulas for the scattered field. Of particular interest is (\ref{eq:26}), that exposes surface wave characteristics to the field near the interface, provided that the numerical distance $k_{01}\uprho\updelta^{2}<1$.


From the analysis of Section \ref{subsec:synopsis}, one might identify that the so-called \emph{Lateral Waves} are not taken into cosideration. Nevertheless,  we still managed to obtain a quite good agreement with the results of numerical methods. As mentioned in\cite{Sautbekov2018}, the ultimate goal is to provide asymptotics, applicable for every possible scenario, not just only for the usual $\upsigma\gg\upomega\upvarepsilon_{0}$ case, considered here. For that purpose, we will insist on the investigation of special function $X\left(k,\alpha\right)\hspace{-1mm}$ and its properties, as well as other special functions that could be used to describe the behavior of the field. Also the obtained asymptotic solutions can be refined with any accuracy and presented in expanded form according to known procedures, to name some of our working group next targets. 

\section*{Acknowledgment}
The authors would like to thank Prof. George J. Fikioris, of the National Technical University of Athens, for making constructive comments towards the preparation of this work.

\bibliographystyle{IEEEtran}
\bibliography{IEEEabrv,Sommerfeld}

\begin{thebibliography}{10}
\providecommand{\url}[1]{#1}
\csname url@samestyle\endcsname
\providecommand{\newblock}{\relax}
\providecommand{\bibinfo}[2]{#2}
\providecommand{\BIBentrySTDinterwordspacing}{\spaceskip=0pt\relax}
\providecommand{\BIBentryALTinterwordstretchfactor}{4}
\providecommand{\BIBentryALTinterwordspacing}{\spaceskip=\fontdimen2\font plus
\BIBentryALTinterwordstretchfactor\fontdimen3\font minus
  \fontdimen4\font\relax}
\providecommand{\BIBforeignlanguage}[2]{{%
\expandafter\ifx\csname l@#1\endcsname\relax
\typeout{** WARNING: IEEEtran.bst: No hyphenation pattern has been}%
\typeout{** loaded for the language `#1'. Using the pattern for}%
\typeout{** the default language instead.}%
\else
\language=\csname l@#1\endcsname
\fi
#2}}
\providecommand{\BIBdecl}{\relax}
\BIBdecl

\bibitem{Sommerfeld1909}
A.~N. Sommerfeld, ``Propagation of {W}aves in {W}ireless {T}elegraphy,''
  \emph{Ann. Phys. (Leipzig)}, vol.~28, pp. 665--737, 1909.

\bibitem{Sommerfeld1926}
------, ``Propagation of {W}aves in {W}ireless {T}elegraphy,'' \emph{Ann. Phys.
  (Leipzig)}, vol.~81, pp. 1135--1153, 1926.

\bibitem{Wait1998}
J.~R. Wait, ``The {A}ncient and {M}odern {H}istory of {EM} {G}round-{W}ave
  {P}ropagation,'' \emph{IEEE Antennas and Propagation Magazine}, vol.~40,
  no.~5, pp. 7--24, Oct. 1998, {DOI}: 10.1109/74.735961.

\bibitem{King1969}
R.~J. King, ``Electromagnetic {W}ave {P}ropagation {O}ver a {C}onstant
  {I}mpedance {P}lane,'' \emph{Radio Science}, vol.~4, no.~3, pp. 255--268,
  Mar. 1969, {DOI}: 10.1029/RS004i003p00255.

\bibitem{Zenneck1907}
J.~Zenneck, ``Propagation of {P}lane {EM} {W}aves along a {P}lane {C}onducting
  {S}urface,'' \emph{Ann. Phys. (Leipzig)}, vol.~23, pp. 846--866, 1907.

\bibitem{Sarkar2012}
T.~K. Sarkar, W.~Dyab, M.~N. Abdallah, M.~Salazar-Palma, M.~V. S.~N. Prasad,
  S.~W. Ting, and S.~Barbin, ``Electromagnetic {M}acro {M}odeling of
  {P}ropagation in {M}obile {W}ireless {C}ommunication: {T}heory and
  {E}xperiment,'' \emph{IEEE Antennas and Propagation Magazine}, vol.~54,
  no.~6, pp. 17--43, Dec. 2012, {DOI}: 10.1109/MAP.2012.6387779.

\bibitem{Bladel2007}
J.~G.~V. Bladel, ``The {S}ommerfeld {D}ipole {P}roblem,'' in
  \emph{Electromagnetic Fields}.\hskip 1em plus 0.5em minus 0.4em\relax
  Hoboken, NJ, USA: J. Wiley and Sons, Inc., 2007, sec. 9.3, pp. 448--452.

\bibitem{Banos1966}
A.~Ba\~nos, \emph{Dipole Radiation in the Presence of a Conducting
  Half-Space}.\hskip 1em plus 0.5em minus 0.4em\relax Oxford, UK: Pergamon
  Press, 1966, pp. 151--158.

\bibitem{Tyras1969}
G.~Tyras, ``Field of a {D}ipole in a {S}tratified {M}edium,'' in
  \emph{Radiation and Propagation of Electromagnetic Waves}.\hskip 1em plus
  0.5em minus 0.4em\relax New York, NY, USA: Accademic Press, Inc., 1969,
  sec.~6, pp. 133--160.

\bibitem{Rahmat1981}
Y.~Rahmat-Samii, R.~Mittra, and P.~Parhami, ``{EVALUATION OF SOMMERFELD
  INTEGRALS FOR LOSSY HALF-SPACE PROBLEMS},'' \emph{Electromagnetics}, vol.~1,
  no.~1, pp. 1--28, 1981, {DOI}: 10.1080/02726348108915122.

\bibitem{Collin2004}
R.~E. Collin, ``Hertzian dipole radiating over a lossy earth or sea: some early
  and late 20th-century controversies,'' \emph{IEEE Antennas and Propagation
  Magazine}, vol.~46, no.~2, pp. 64--79, Apr. 2004, {DOI}:
  10.1109/MAP.2004.1305535.

\bibitem{Michalski1985}
K.~A. Michalski, ``On the efficient evaluation of integral arising in the
  sommerfeld halfspace problem,'' \emph{IEE Proceedings H - Microwaves,
  Antennas and Propagation}, vol. 132, no.~5, pp. 312--318, Aug. 1985, {DOI}:
  10.1049/ip-h-2.1985.0056.

\bibitem{Pelosi2010}
G.~Pelosi and J.~L. Volakis, ``On the {C}entennial of {S}ommerfeld's {S}olution
  to the {P}roblem of {D}ipole {R}adiation {O}ver an {I}mperfectly {C}onducting
  {H}alf {S}pace,'' \emph{IEEE Antennas and Propagation Magazine}, vol.~52,
  no.~3, pp. 198--201, Jun. 2010, {DOI}: 10.1109/MAP.2010.5586629.

\bibitem{Norton1936}
K.~A. Norton, ``The {P}ropagation of {R}adio {W}aves over the {S}urface of the
  {E}arth and in the {U}pper {A}tmosphere,'' \emph{Proceedings of the Institute
  of Radio Engineers}, vol.~24, no.~10, pp. 1367--1387, Oct. 1936, {DOI}:
  10.1109/JRPROC.1936.227360.

\bibitem{Norton1937}
------, ``The {P}ropagation of {R}adio {W}aves over the {S}urface of the
  {E}arth and in the {U}pper {A}tmosphere,'' \emph{Proceedings of the Institute
  of Radio Engineers}, vol.~25, no.~9, pp. 1203--1236, Sep. 1937, {DOI}:
  10.1109/JRPROC.1937.228544.

\bibitem{CEMA2010}
\BIBentryALTinterwordspacing
S.~S. Sautbekov, R.~N. Kasimkhanova, and P.~V. Frangos, ``{M}odified {S}olution
  of {S}ommerfelds's {P}roblem,'' in \emph{Proc. CEMA’10}, Athens, Greece,
  Oct. 07 -- 09, 2010, pp. 5--8. [Online]. Available:
  \url{http://rcvt.tu-sofia.bg/CEMA/proceedings/CEMA\_2010\_proc.pdf}
\BIBentrySTDinterwordspacing

\bibitem{Sautbekov2011}
\BIBentryALTinterwordspacing
S.~Sautbekov, ``The {G}eneralized {S}olutions of a {S}ystem of {M}axwell's
  {E}quations for the {U}niaxial {A}nisotropic {M}edia,'' in
  \emph{Electromagnetic Waves Propagation in Complex Matter}.\hskip 1em plus
  0.5em minus 0.4em\relax London, UK: IntechOpen Limited, 2011, ch.~1, pp.
  1--24. [Online]. Available: \url{https://doi.org/10.5772/16886}
\BIBentrySTDinterwordspacing

\bibitem{Ioannidi2014}
K.~Ioannidi, {\relax Ch}.~Christakis, S.~Sautbekov, P.~Frangos, and S.~K.
  Atanov, ``The {R}adiation {P}roblem from a {V}ertical {H}ertzian {D}ipole
  {A}ntenna above {F}lat and {L}ossy {G}round: {N}ovel {F}ormulation in the
  {S}pectral {D}omain with {C}losed -- {F}orm {A}nalytical {S}olution in the
  {H}igh {F}requency {R}egime,'' \emph{International Journal of Antennas and
  Propagation (IJAP), Special Issue on Propagation of Electromagnetic Waves in
  Terrestrial Environment for Applications in Wireless
  Telecommunications\textnormal{, Hindawi Ed. Co.}}, vol. 2014, {A}rticle ID
  989348, 9 Pages, Aug. 2014, {DOI}: 10.1155/2014/989348.

\bibitem{Balanis1997}
C.~A. Balanis, ``Method of {S}tationary {P}hase,'' in \emph{Antenna Theory:
  Analysis and Design}, 2nd~ed.\hskip 1em plus 0.5em minus 0.4em\relax New
  York, NY, USA: J. Wiley and Sons, Inc., 1997, {A}ppendix VIII, pp. 922--927.

\bibitem{Moschovitis2010}
C.~G. Moschovitis, K.~T. Karakatselos, E.~G. Papkelis, H.~T. Anastassiu, I.~C.
  Ouranos, A.~Tzoulis, and P.~V. Frangos, ``Scattering of {E}lectromagnetic
  {W}aves {F}rom a {R}ectangular {P}late {U}sing an {E}nhanced {S}tationary
  {P}hase {M}ethod {A}pproximation,'' \emph{IEEE Transactions on Antennas and
  Propagation}, vol.~58, no.~1, pp. 233--238, Jan 2010, {DOI}:
  10.1109/TAP.2009.2024015.

\bibitem{Moschovitis2010PIER}
C.~G. Moschovitis, H.~T. Anastassiu, and P.~V. Frangos, ``Scattering of
  {E}lectromagnetic {W}aves from a {R}ectangular {P}late {U}sing an {E}xtended
  {S}tationary {P}hase {M}ethod {B}ased on {F}resnel {F}unctions {(SPM-F)},''
  \emph{Progress In Electromagnetics Research}, vol. 107, pp. 63--99, 2010,
  {DOI}: 10.2528/PIER10040104.

\bibitem{Bourgiotis2014}
\BIBentryALTinterwordspacing
S.~Bourgiotis, K.~Ioannidi, {\relax Ch}.~Christakis, S.~Sautbekov, and
  P.~Frangos, ``The {R}adiation {P}roblem from a {V}ertical {S}hort {D}ipole
  {A}ntenna {A}bove {F}lat and {L}ossy {G}round: {N}ovel {F}ormulation in the
  {S}pectral {D}omain with {N}umerical {S}olution and {C}losed-{F}orm
  {A}nalytical {S}olution in the {H}igh {F}requency {R}egime,'' in \emph{Proc.
  CEMA’14}, Sofia, Bulgaria, Oct. 16 -- 18, 2014, pp. 12--18. [Online].
  Available:
  \url{http://rcvt.tu-sofia.bg/CEMA/proceedings/CEMA\_2014\_proc.pdf}
\BIBentrySTDinterwordspacing

\bibitem{Bourgiotis2015}
S.~Bourgiotis, A.~Chrysostomou, K.~Ioannidi, S.~Sautbekov, and P.~Frangos,
  ``Radiation of a {V}ertical {D}ipole over {F}lat and {L}ossy {G}round using
  the {S}pectral {D}omain {A}pproach: {C}omparison of {S}tationary {P}hase
  {M}ethod {A}nalytical {S}olution with {N}umerical {I}ntegration {R}esults,''
  \emph{Electronics and Electrical Engineering}, vol.~21, no.~3, pp. 38--41,
  Jun. 2015, {DOI}: 10.5755/j01.eee.21.3.10268.

\bibitem{Chrysostomou2016}
A.~Chrysostomou, S.~Bourgiotis, S.~Sautbekov, K.~Ioannidi, and P.~V. Frangos,
  ``Radiation of a {V}ertical {D}ipole {A}ntenna over {F}lat and {L}ossy
  {G}round: {A}ccurate {E}lectromagnetic {F}ield {C}alculation using the
  {S}pectral {D}omain {A}pproach along with {R}edefined {I}ntegral
  {R}epresentations and corresponding {N}ovel {A}nalytical {S}olution,''
  \emph{Electronics and Electrical Engineering}, vol.~22, no.~2, pp. 54--61,
  Apr. 2016, {DOI}: 10.5755/j01.eie.22.2.14592.

\bibitem{Sautbekov2018}
S.~Sautbekov, S.~Bourgiotis, A.~Chrysostomou, and P.~Frangos, ``A {N}ovel
  {A}symptotic {S}olution to the {S}ommerfeld {R}adiation {P}roblem: {A}nalytic
  {F}ield {E}xpressions and the {E}mergence of the {S}urface {W}aves,''
  \emph{Progress in Electromagnetic Research}, vol.~64, pp. 9--22, 2018, {DOI}:
  10.2528/PIERM17082806.

\bibitem{Weinstein1969}
L.~A. Weinstein, \emph{The Theory of Diffraction and the Factorization
  Method}.\hskip 1em plus 0.5em minus 0.4em\relax Boulder, CO, USA: Golem
  Press, 1969.

\bibitem{Fock1945}
V.~A. Fock, ``Diffraction of {R}adio {W}aves {A}round the {E}arth's
  {S}urface,'' \emph{Jour. Phys. U.S.S.R.}, vol.~9, pp. 256--266, 1945.

\bibitem{Fock1946}
------, ``The {F}ield of a {P}lane {W}ave {N}ear the {S}urface of a
  {C}onducting {B}ody,'' \emph{Jour. Phys. U.S.S.R.}, vol.~10, pp. 399--409,
  1946.

\bibitem{Leontovic1946}
M.~A. Leontovic and V.~A. Fock, ``Propagation of {E}lectromagnetic {W}aves
  {A}long the {E}arth's {S}urface,'' \emph{Jour. Phys. U.S.S.R.}, vol.~10, pp.
  13--24, 1946.

\bibitem{Pedlosky2003}
J.~Pedlosky, \emph{Waves in the Ocean and Atmosphere: Introduction to Wave
  Dynamics}.\hskip 1em plus 0.5em minus 0.4em\relax Berling - Heidelberg,
  Germany: Springer, 2003.

\bibitem{Sautbekov2010}
S.~Sautbekov, ``Factorization {M}ethod for {F}inite {F}ine {S}tructures,''
  \emph{Progress in Electromagnetic Research B}, vol.~25, pp. 1--21, 2010,
  {DOI}: 10.2528/PIERB10071801.

\bibitem{DLMF}
\BIBentryALTinterwordspacing
``{NIST Digital Library of Mathematical Functions},'' http://dlmf.nist.gov/,
  Release 1.0.23 of 2019-06-15, f.~W.~J. Olver, A.~B. {Olde Daalhuis}, D.~W.
  Lozier, B.~I. Schneider, R.~F. Boisvert, C.~W. Clark, B.~R. Miller and B.~V.
  Saunders, eds. ch. 7. [Online]. Available: \url{http://dlmf.nist.gov/7}
\BIBentrySTDinterwordspacing

\bibitem{Bourgiotis2018}
\BIBentryALTinterwordspacing
S.~Bourgiotis, L.~Dimopoulos, {\relax Th}.~Lymperopoulos, A.~Chrysostomou,
  S.~Sautbekov, and P.~Frangos, ``Radiation of {V}ertical {D}ipole {A}ntennas
  {O}ver {F}lat and {L}ossy {T}errain: {A} {N}ovel {E}fficient {M}ethod for the
  {A}ccurate {N}umerical {C}alculation of the {S}ommerfeld {I}ntegrals in the
  {S}pectral {D}omain,'' in \emph{Proc. CEMA’18}, Sofia, Bulgaria, Oct. 18 --
  20, 2018, pp. 6--11. [Online]. Available:
  \url{http://rcvt.tu-sofia.bg/CEMA/proceedings/CEMA\_2018\_proc.pdf}
\BIBentrySTDinterwordspacing

\bibitem{Fikioris2013}
G.~Fikioris, I.~Tatsoglou, and O.~N. Bakas, ``On the {C}onvergence/{D}ivergence
  of {D}efinite {I}ntegrals,'' in \emph{Selected Asymptotic Methods with
  Applications to Electromagnetics and Antennas}, C.~A. Balanis, Ed.\hskip 1em
  plus 0.5em minus 0.4em\relax San Rafael, CA, USA: Morgan \& Claypool, 2013,
  {A}ppendix~B, pp. 173--178.

\bibitem{Fikioris1982}
J.~Fikioris, \emph{Introduction to Antenna Theory and Propagation of
  Electromagnetic Waves}.\hskip 1em plus 0.5em minus 0.4em\relax Athens,
  Greece: National Technical University of Athens, 1982, (in Greek).

\bibitem{Arfken2005}
G.~B. Arfken and H.~J. Weber, ``Hankel {F}unctions,'' in \emph{Mathematical
  Methods for Physicists}, 6th~ed.\hskip 1em plus 0.5em minus 0.4em\relax
  Burlington, MA, USA: Elsevier Inc., 2005, ch. 11.4, pp. 707--711.

\bibitem{Olver1964}
F.~W.~J. Olver, ``Bessel {F}unctions of {I}nteger {O}rder,'' in \emph{Handbook
  of Mathematical Functions with Formulas, Graphs, and Mathematical Tables},
  ninth dover printing~ed., M.~Abramowitz and I.~A. Stegun, Eds.\hskip 1em plus
  0.5em minus 0.4em\relax New York, NY, USA: Dover, 1964, ch.~9, p. 361.

\bibitem{IEEEterms}
``{IEEE} {S}tandard {D}efinitions of {T}erms for {R}adio {W}aves
  {P}ropagation,'' \emph{IEEE Std 211-2018 (Revision of IEEE Std 211-1997)},
  pp. 1--57, Feb 2019.

\bibitem{Thomas2010}
G.~Thomas, M.~Weir, and J.~Hass, ``Numerical {I}ntegration,'' in \emph{Thomas'
  Calculus}, 12th~ed.\hskip 1em plus 0.5em minus 0.4em\relax New York, NY, USA:
  Addison-Wesley, 2010, ch. 8.6, pp. 486--493.

\bibitem{McKeeman1962}
W.~M. McKeeman, ``Algorithm 145: {A}daptive {N}umerical {I}ntegration by
  {S}impson's {R}ule,'' \emph{Commun. ACM}, vol.~5, no.~12, p. 604, Dec. 1962,
  {DOI}: 10.1145/355580.369102.

\bibitem{Hillstrom1970}
K.~E. Hillstrom, ``Comparison of several adaptive {N}ewton-{C}otes quadrature
  routines in evaluating definite integrals with peaked integrands,''
  \emph{Commun. ACM}, vol.~13, no.~6, pp. 362--365, Jun. 1970, {DOI}:
  10.1145/362384.362499.

\bibitem{Zaghloul2012}
M.~R. Zaghloul and A.~N. Ali, ``Algorithm 916: Computing the {F}addeyeva and
  {V}oigt functions,'' \emph{ACM Trans. Math. Softw.}, vol.~38, no.~2, pp.
  15:1--15:22, Jan. 2012, {DOI}: 10.1145/2049673.2049679.

\bibitem{Zaghloul2017}
M.~R. Zaghloul, ``Algorithm 985: {S}imple, efficient, and relatively accurate
  approximation for the evaluation of the {F}addeyeva function,'' \emph{ACM
  Trans. Math. Softw.}, vol.~44, no.~2, pp. 22:1--22:9, Aug. 2017,
  10.1145/3119904.

\bibitem{Bender1999}
C.~M. Bender and S.~A. Orszag, ``Asymptotic {E}xpansion of {I}ntegrals,'' in
  \emph{Mathematical Methods for Scientists and Engineers; Asymptotic Methods
  and Perturbation theory}.\hskip 1em plus 0.5em minus 0.4em\relax New York,
  NY, USA: Springer, 1999, ch.~6, pp. 247--316.

\end{thebibliography}

\end{document}